\documentclass[sigconf,authorversion]{acmart}
\usepackage{colortbl}
\usepackage{amsmath}
\usepackage{graphicx,calc}
\usepackage{tabularx}
\usepackage{xcolor}
\usepackage{enumitem} 
\usepackage{graphicx} 
\usepackage{lipsum} 
\newlength\myheight
\newlength\mydepth
\settototalheight\myheight{Xygp}
\usepackage{booktabs} 
\usepackage{array} 
\usepackage{subcaption}
\usepackage{fontawesome5}
\usepackage{dblfloatfix}

\newcommand{\rev}[1]{{\color{black}#1}}

\AtBeginDocument{%
  }

\copyrightyear{2026}
\acmYear{2026}
\setcopyright{cc}
\setcctype{by-nc-nd}
\acmConference[CHI '26]{Proceedings of the 2026 CHI Conference on Human Factors in Computing Systems}{April 13--17, 2026}{Barcelona, Spain}
\acmBooktitle{Proceedings of the 2026 CHI Conference on Human Factors in Computing Systems (CHI '26), April 13--17, 2026, Barcelona, Spain}
\acmPrice{}
\acmDOI{10.1145/3772318.3790572}
\acmISBN{979-8-4007-2278-3/2026/04}

\begin{document}
\newcommand{\sysname}[0]{\textsc{DiaryPlay}}
\newcommand{\daehyun}[1]{{\color{teal}\bf{DH: #1}\normalfont}}
\newcommand{\jiangnan}[1]{{\color{red}\bf{JN: #1}\normalfont}}
\newcommand{\haeseul}[1]{{\color{purple}\bf{HS: #1}\normalfont}}
\newcommand{\gosu}[1]{{\color{orange}\bf{GS: #1}\normalfont}}
\newcommand{\kp}[1]{{\color{orange}\bf{KP: #1}\normalfont}}
\newcommand{\juho}[1]{{\color{blue}\bf{JH: #1}\normalfont}}

\newcommand{\revision}[1]{\textcolor{purple}{#1}}
\newcommand{\revisedh}[1]{\textcolor{red}{#1}}

\title[\sysname{}]{\sysname{}: AI-Assisted Creation of Interactive Story Vignettes for Everyday Storytelling}

\author{Jiangnan Xu}
\email{jiangnan.xu@tuni.fi}
\orcid{0000-0002-8932-8082}
\affiliation{%
  \institution{Rochester Institute of Technology}
  \city{Rochester}
  \state{NY}
  \country{USA}
  }
\affiliation{%
  \institution{Tampere University}
  \city{Tampere}
  \country{Finland}
}

\author{Haeseul Cha}
\email{jjchs1@kaist.ac.kr}
\orcid{0009-0009-8512-278X}
\affiliation{%
  \institution{KAIST}
  \city{Daejeon}
  \country{Korea}
}

\author{Gosu Choi}
\email{coregosu1227@gmail.com}
\orcid{0000-0002-1991-967X}
\affiliation{%
  \institution{KAIST}
  \city{Daejeon}
  \country{Korea}
}

\author{Gyu-cheol Lee}
\email{gc.lee@kt.com}
\orcid{0000-0002-7814-355X}
\affiliation{%
  \institution{Korea Telecom}
  \city{Seoul}
  \country{Korea}
}

\author{Yeo-Jin Yoon}
\email{yjin.yun@kt.com}
\orcid{0000-0002-4101-2011}
\affiliation{%
  \institution{Korea Telecom}
  \city{Seoul}
  \country{Korea}
}

\author{Zucheul Lee}
\email{polelee@kt.com}
\orcid{0000-0002-7814-355X}
\affiliation{%
  \institution{Korea Telecom}
  \city{Seoul}
  \country{Korea}
}

\author{Konstantinos Papangelis}
\email{kxpigm@g.rit.edu}
\orcid{0000-0002-7184-2753}
\affiliation{%
  \institution{Rochester Institute of Technology}
  \city{Rochester}
  \state{NY}
  \country{USA}
}

\author{Dae Hyun Kim}
\authornote{Co-corresponding authors}
\email{dhkim16@yonsei.ac.kr}
\orcid{0000-0002-8657-9986}
\affiliation{%
  \institution{Yonsei University}
  \city{Seoul}
  \country{Korea}
}
\affiliation{%
  \institution{POSTECH}
  \city{Pohang}
  \country{Korea}
}

\author{Juho Kim}
\authornotemark[1]
\email{juho@juhokim.com}
\orcid{0000-0001-6348-4127}
\affiliation{%
  \institution{KAIST}
  \city{Daejeon}
  \country{Korea}
}
\affiliation{%
  \institution{SkillBench}
  \city{Santa Barbara, CA}
  \country{USA}
}

\renewcommand{\shortauthors}{Xu et al.}

\begin{abstract}
An interactive vignette is a visual storytelling medium that lets the audience role-play a character and interact with non-player characters (NPCs) and the digital environment. 
Yet, the authoring complexity of interactive vignettes has obstructed their adoption in everyday storytelling, which builds on immediacy. 
We introduce \sysname{}, an AI-assisted authoring system that generates interactive vignettes from text stories. 
The \textit{Authoring Component} visually elicits three core elements (environment, characters, events) through automation and author refinement. 
The \textit{Viewing Component} delivers an interactive story to the audience using an LLM-powered \textit{Controlled Divergence Module}, which allows divergent player and NPC behaviors within the boundaries defined by the author's intended story. 
A technical evaluation shows that the Controlled Divergence module generates believable NPC activities based on both character persona and storyline. 
A user study demonstrates that \sysname{} enables low-effort authoring of interactive vignettes for everyday storytelling while providing engaging viewing experiences and conveying the core story message.

\end{abstract}

\ccsdesc[500]{Human-centered computing~Empirical studies in HCI}
\ccsdesc[500]{Human-centered computing~Interactive systems and tools}

\keywords{Interactive Storytelling, Everyday Storytelling, Interactive Vignette, Role-Playing, Authoring System, Human-AI Collaboration, Large Language Model}
\begin{teaserfigure}
  \centering
  \includegraphics[width=0.84\textwidth]{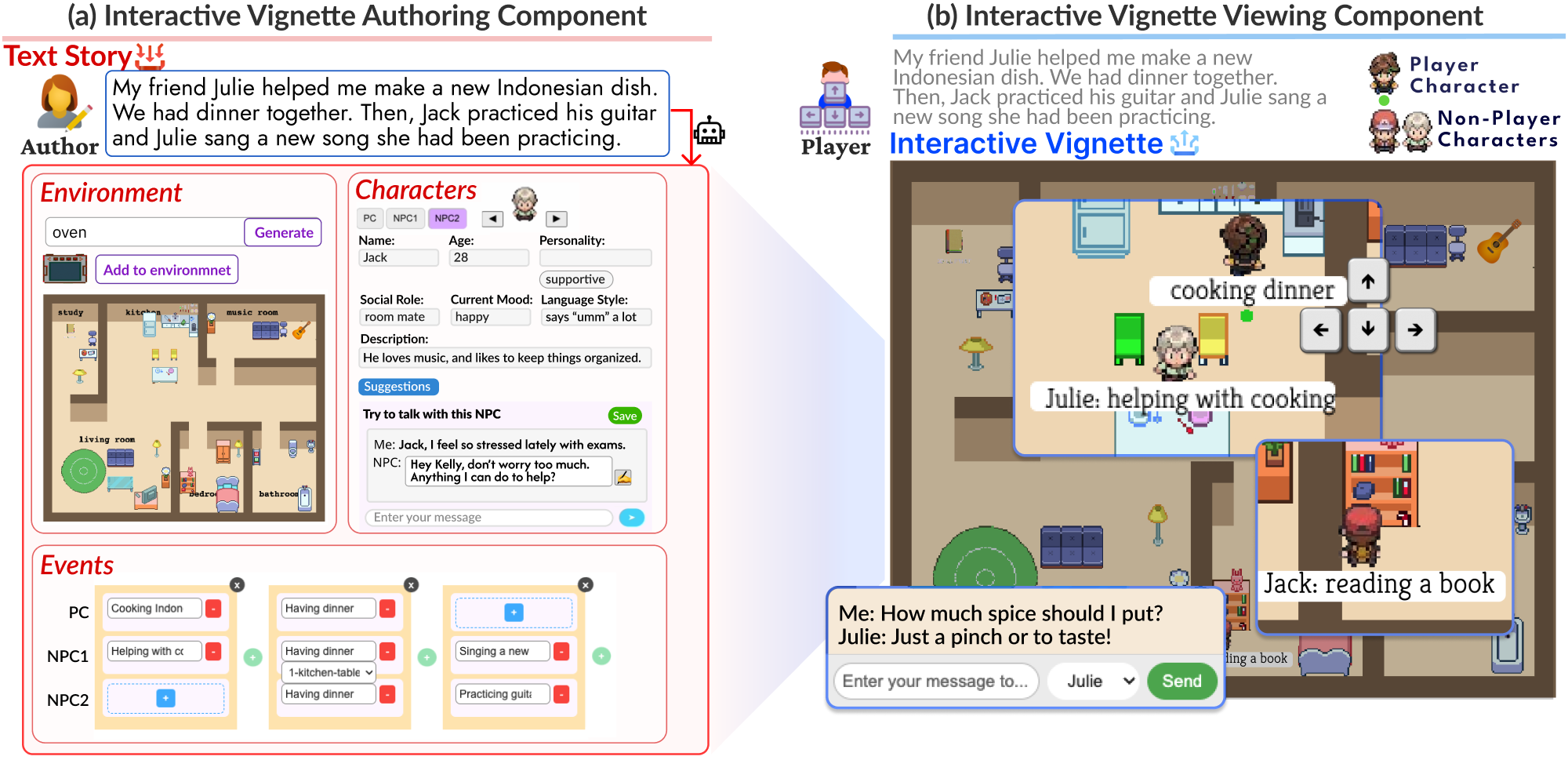}
  \caption{\sysname{} assists everyday storytellers in creating interactive vignettes. It contains two components: (a) The Interactive Vignette Authoring Component, which guides the everyday storyteller (author) to complete and refine the interactive vignette elements through surfacing the gap between what is captured by the natural language story input and the information requirements for an interactive vignette, (b) The Interactive Vignette Viewing Component, which enables the audience (player) to role-play the main character (player character; PC) to follow the author-defined storyline or engage in divergent activities within the environment, and plan non-player characters' (NPCs) behaviors in response to the player interactions.}
  \Description{\sysname{} empowers casual authors to craft interactive vignettes by taking everyday stories in text form as input and utilizing LLM support to facilitate the authoring process. The completed interactive vignette allows the audience to immerse themselves in the author's narrative by assuming the role of the author as the player character and interacting with other non-player characters and objects within a virtual environment.}
  \label{fig:teaser}
\end{teaserfigure}


\maketitle

\section{Introduction}
Everyday storytelling is an intrinsic part of human nature~\cite{kelliher2007everyday}.
People not only tell stories about daily moments verbally to those around them, but also share everyday stories digitally on social media platforms (e.g., X~\cite{x}, Instagram~\cite{instagram}) by posting text, photos, or short videos~\cite{goody2006oral,choo2020digital}.
Regardless of the medium, everyday storytelling is rooted in \textit{immediacy}~\cite{perez2016social}, characterized by minimal curation and editing~\cite{villaespesa2020ephemeral}, which enables in-the-moment sharing when the stories remain timely and relevant.
Unlike professional forms of storytelling such as novels or films, everyday storytelling is produced by the general public and occupies the lightweight, casual end of the storytelling spectrum~\cite{kim2024diarymate}.



Although \textit{interactive vignettes}---a form of visual storytelling in which the \rev{audience, acting as a player,} controls a character and participates by interacting with characters and objects (e.g., interactive drama~\cite{facade2005,hollywoodreporter2018bandersnatch}, role-playing video games~\cite{eldenring2022,thesims})---have gained popularity as an engaging digital medium~\cite{gordon2011playing,konstantopoulou2018designing,spierling2006learning}, they have not yet been adopted for everyday storytelling. 
\rev{Interactive vignettes possess two core qualities that can aid immersion of the audience into the narrative: (1) \textit{interactivity} and (2) \textit{first-hand participatory experiences}; these core qualities \textit{together} give interactive vignettes a potential to enrich everyday storytelling alongside familiar social-media formats such as text, images, and video.}
However, existing tools for authoring interactive vignettes (e.g., RPGMaker~\cite{rpgmaker}, Unity~\cite{unity}, and Roblox Studio~\cite{roblox}) are designed for producing polished works, with development workflows that often span months or years~\cite{owens2011social}. 
\rev{Interactive vignette authoring often requires authors to curate interactive story scenes~\cite{fime2024automatic} and craft narrative branches that diverge into multiple situations~\cite{koenitz2015Practicalities, spierling2012implicit}}, which is too technically demanding and time-intensive to accommodate the immediacy required in everyday storytelling.

A formative study (Section~\ref{sec:formative}) with 13 active social media users confirmed the desire to adopt interactive vignettes for everyday storytelling. 
Participants valued interactive vignettes, \rev{especially for everyday stories centered on events}, because they can let the audience role-play the author's perspective to unfold events and foster a sense of resonance with the author's personal experience, despite the authoring challenges that hinder immediacy required for everyday storytelling.
We further derive three design goals for interactive vignette authoring tools for everyday storytelling: (1) enable a lightweight authoring process that matches the effort of everyday storytelling practices, (2) capture and convey the author's key message in the everyday story, 
and (3) deliver an engaging viewing experience with situation-aware non-player characters (NPCs) in a branching storyline.

Based on the design goals, we introduce \sysname{} (Figure~\ref{fig:teaser}), an AI-assisted authoring system for creating interactive vignettes tailored to the lightweight and casual, everyday storytelling context.
At a high level, the design of \sysname{} explores the balance between human authoring and AI generation in narrative authoring contexts; users can direct the narrative authoring process of interactive vignettes without explicitly specifying all low-level details.

\sysname{} takes a natural language text story as input and outputs an interactive vignette, where the player controls the main character (player character; PC) to perform activities by interacting with objects in the environment, and watches other characters (NPCs) perform activities and engage in conversations with NPCs.
The authoring system comprises two main components: (1) the \textit{Interactive Vignette Authoring Component}, which supports a lightweight authoring process, through automation and refinement, visually guides the author in specifying the three core elements of an interactive vignette: environment, characters, and events; and
(2) the \textit{Interactive Vignette Viewing Component}, which delivers an interactive experience to the player, by enlivening NPCs reacting to player activities and dialogues.
The Interactive Vignette Viewing Component relies on the \textit{Controlled Divergence (CD) Module}, an automatic LLM-based module that uses the branch-and-bottleneck~\cite{rezk2022beyond} storytelling structure to dynamically plan NPC behaviors and allow divergent activities while preserving the author's main messages.
With the CD Module, interactive vignettes can branch adaptively without extra author effort on multi-branching. 

A preliminary technical evaluation (Section~\ref{ter}) of 16 participants shows that the CD Module generates believable NPC activities based on the given character persona and storyline.
To investigate how \sysname{} supports everyday storytelling, we conducted a user study with 16 participants (Section~\ref{sec:userstudy}). 
The authoring experience met participants' expectations for lightweight effort; they were able to complete interactive vignette creation within 20 minutes (see resulting examples in Figure~\ref{fig:result}), and \sysname{} provided effective assistance in building interactive vignette elements. 
Participants valued the moments when players diverged from the main storyline, bringing in tiny yet meaningful pieces that enriched but did not derail their everyday story.
Meanwhile, participants described the viewing experience as engaging and immersive, and they were able to understand the author's main story content despite divergent interactions.

In summary, this paper makes the following contributions:
\begin{itemize}[leftmargin=*,topsep=0pt]
    \item \sysname{}, an end-to-end AI-assisted interactive vignette authoring system for everyday storytelling;
    \item the Controlled Divergence (CD) Module, an LLM-powered module that allows divergent player and NPC behaviors within the boundaries defined by the author's intended story; 
    \item results of a technical evaluation showing that the CD Module generates believable NPC activities based on both character persona and storyline; and
    \item empirical findings from a user study, demonstrating that \sysname{} enables low-effort authoring of interactive vignettes for everyday storytelling while providing engaging viewing experiences and conveying the core story message.
\end{itemize}

\section{Related Work}
Our work is related to three main areas of research: (1) interactive vignette and games authoring, (2) building multi-branch narratives, and (3) LLM-powered believable characters in storytelling.

\subsection{Interactive Vignette and Games Authoring}
In interactive digital storytelling, role-playing~\cite{peinado2004transferring} is a widely used approach that allows players to control a character and interact with other characters and objects within the story.
Role-playing fosters deep immersion and empathy~\cite{gordon2011playing, konstantopoulou2018designing}, resulting in memorable storytelling experiences~\cite{spierling2006learning}.
Our work focuses on role-playing through visual storytelling, excluding purely text-based interactive narratives~\cite{urbanek2019learning, xi2021kuileixi}.
We adopt the term \textit{interactive vignette} to describe a genre of visual storytelling media that integrates role-playing within an interactive environment (e.g., interactive drama~\cite{verdugo2011interactive, hales2015interactive, hollywoodreporter2018bandersnatch, facade2005} and role-playing games (RPGs)~\cite{minecraft, thesims, eldenring2022}).
Our work adopts the framework proposed by Zhao et al.~\cite{Zhao2024}, which defines an interactive vignette through three core elements: environment, characters, and events.

Traditionally, creating interactive vignettes has been considered resource-intensive and time-consuming~\cite{kumaran2024narrativegenie, qin2024charactermeet, authoringIssue, fime2024automatic}.
Authoring tools like Unity~\cite{unity} and RPGMaker~\cite{rpgmaker} require authors to conceptualize assets, design or source visual representations, manage interactions, and carefully script narrative events.
\rev{In game design, a growing body of casual creator tools~\cite{compton2015casual} supports lightweight development for non-professional users. Bitsy~\cite{bitsy} offers a minimalist 2D tilemap editor for small pixel-art narrative scenes, PuzzleScript~\cite{puzzlescript} provides a domain-specific rule language for rapidly prototyping puzzle games, and Downpour~\cite{downpourTool} enables collaging text, images, and drawings into simple branching narratives via clickable buttons. While these tools are not designed for everyday storytelling, they reflect a broader trend of empowering the public to create interactive media. To further lower authoring barriers, research has explored workflows where authors express high-level intent and systems generate interactive artifacts: Germinate~\cite{Kreminski2020germinate} and Game-O-Matic~\cite{Treanor2012game-o-matic} use generative models to create games from conceptual inputs, Wevva~\cite{Powley2021Wevva} lets users adjust parameters during gameplay, and RosebudAI~\cite{rosebudAI} uses generative AI to produce game elements from text prompts. However, unlike tools oriented toward game design, \sysname{} focuses on everyday storytelling rather than strategy, mechanics, or rules.
}

Although user-generated storytelling content (e.g., vlogs, blogs) is widespread, interactive vignette creation remains uncommon for everyday storytellers. Recent advances in generative AI, especially LLMs, have streamlined the creation of interactive storytelling elements such as role-play chatbots~\cite{Zhao2024, kumaran2023scenecraft,sun2022bringing}, interactive objects~\cite{ye2025mographgpt}, scenes~\cite{bruce2024genie}, autonomous characters~\cite{park2023generative}, and narrative beats~\cite{kumaran2024narrativegenie}. AI-assisted tools can now generate content directly from natural-language inputs~\cite{fan2025words}, reducing traditional authoring complexity, but may also diminish author agency~\cite{fisher2023centering}. 
Human-in-the-loop~\cite{zanzotto2019human} systems such as PatchView~\cite{chung2024patchview} and FairyTailor~\cite{bensaid2021fairytailor} address this by pairing AI automation with human modification. Building on this collaborative approach~\cite{Wu2021AI}, our work uses LLM-driven generation to reduce the authoring burden of interactive vignettes while preserving author agency throughout the process.

\subsection{Building Multi-Branch Interactive Narratives}
\rev{
Interactive vignettes foster audience engagement by enabling exploration and influence over narrative progression~\cite{kumaran2024narrativegenie}, but they also require narratives to respond dynamically to player interactions while maintaining coherence~\cite{aylett2011research, riedl2013interactive}. Because authors are absent during play, they must predefine structures that anticipate diverse interactions, often through multi-branch storytelling. Kumaran et al.~\cite{kumaran2024narrativegenie} describe two main approaches: representing narratives as story graphs, where states are connected by causal edges and a runtime component manages progression, and enabling the runtime to revise narrative elements dynamically as the story unfolds. In the story-graph approach, tools such as StoryTec~\cite{gobel2008storytec} require authors to specify branch conditions and actions, and art-E-fact~\cite{linaza2004authoring} similarly relies on constructing narrative graphs, while AI-assisted systems like GENEVA~\cite{leandro2024geneva} and Spindle~\cite{calderwood2022spinning} support authors in building these branching structures. In the dynamic-revision approach, StoryVerse~\cite{wang2024storyverse} combines LLM-driven character simulation with an abstract-act authorial structure to support emergent narratives~\cite{aylett2000emergent}, though it limits authors to specifying high-level conflicts or turning points rather than complete stories. Likewise, Façade~\cite{mateas2003faccade} provides AI assistance by dynamically selecting and sequencing pre-authored branches (behavior trees) in response to player interactions to maintain a coherent storyline.
}

However, everyday storytellers, who are often non-expert storytellers, are typically accustomed to single-branch storytelling, and the requirement to conceptualize multi-branch narratives can contradict their natural storytelling habits, making the authoring process challenging~\cite{koenitz2015Practicalities, spierling2012implicit}.
To address this, we aim to develop an implicit approach that transforms single-branch stories into interactive narratives without requiring authors to manually craft multiple branching paths~\cite{spierling2012implicit}.
NarrativeGenie~\cite{kumaran2024narrativegenie} represents an initial attempt.
It creates a partially ordered sequence of events based on an author's high-level story description and uses an LLM-based pipeline to select and reorder events in response to player interactions.
However, in everyday storytelling, event sequence is crucial in preserving the author-intended temporal flow~\cite{labov1997narrative}.
Hence, we aim to realize branching narrative generation during story runtime that maintains the authored story's progression while enabling emergent narratives~\cite{aylett2000emergent} to react to dynamic player interactions.


\subsection{LLM-powered Believable Characters in Interactive Digital Storytelling}
Virtual characters (e.g., AI agents~\cite{park2023generative}) are designed to interact with users and simulate real-life behaviors in interactive digital storytelling. 
To create engaging and immersive experiences, \textit{believability} has been set as a central design goal~\cite{park2023generative}, meaning characters should exhibit lifelike behaviors, make decisions autonomously, and respond naturally to evolving scenarios~\cite{thomas1995illusion, bogdanovych2016makes}. 
\rev{A well-defined persona is essential for believable characters~\cite{loyall1997believable, lee2012you, aljammaz2023towards}. 
Beyond traditional attribute-based specification, recent work explores conversation-based persona construction in which authors build characters through dialogue with an LLM to better finetune its persona~\cite{qin2024charactermeet}. 
Commercial tools (e.g., CharacterAI~\cite{characterai}) similarly let users define descriptions, greetings, example dialogues, and personality tags, with iterative refinement through response ratings.
}
~\rev{Beyond initial character construction, sustaining believable interactions requires a character architecture that maintains persona consistency; the generative agent framework~\cite{park2023generative} addresses this through memory-based perception, storage, and retrieval to support adaptive yet coherent behavior. 
Beyond making character behaviors consistent with their persons, our work explores characters that remain believable while \textit{also} aligning with an author's intended storyline~\cite{lu2025whatelse,ware2021sabre}.
}

While existing work primarily focuses on the importance of persona consistency~\cite{xiao2023far}, in storytelling, we argue that believable character behaviors should also ensure the narrative's logical coherence~\cite{riedl2010narrative}.
However, persona consistency and narrative coherence might conflict in a narrative planning~\cite{ware2014glaive, riedl2010narrative}. 
For instance, if a character is portrayed as a person who loves traveling and is set in the storyline to ``work hard for a final exam'' as a future activity, their behavior becomes less believable if they are shown engaging in a ``go on a road trip.'' 
Because it might disrupt the logical flow of the narrative, even if their activity remains consistent with their persona.
To address this issue, our work proposes a character activity planning approach that considers both the character's persona consistency as well as narrative coherence.
\section{Formative Study}
\label{sec:formative}
To understand people's motivations for authoring and viewing interactive vignettes in everyday storytelling, and to derive design goals for an authoring system to support this practice, we conducted an IRB-approved formative study.

\subsection{Procedure}
We recruited 13 participants (6 females, 7 males, aged 19 to 35) through online communities within KAIST.
To capture potential users of interactive vignette systems, we recruited active social media users who regularly engage with everyday storytelling, creating and consuming digital storytelling (e.g., social media posts, blogs, vlogs) at least once a week.
\rev{
Participants were from various academic backgrounds and reported various levels of frequency of interactive-vignette consumption (e.g., play RPGs, watch interactive drama), including eight participants who play RPGs at least three times per week.
We list detailed participants' background information in Table~\ref{tab:pilotparticipants} at Appendix~\ref{app:participant}.}
For each participant, two researchers conducted the formative study in person, with one researcher leading the study and the other taking notes.
The study lasted roughly 1.5 hours, and each participant received 22,000 KRW ($\approx 17$ USD) as compensation. 

\begin{figure}[]
    \centering
    \includegraphics[width=\linewidth]{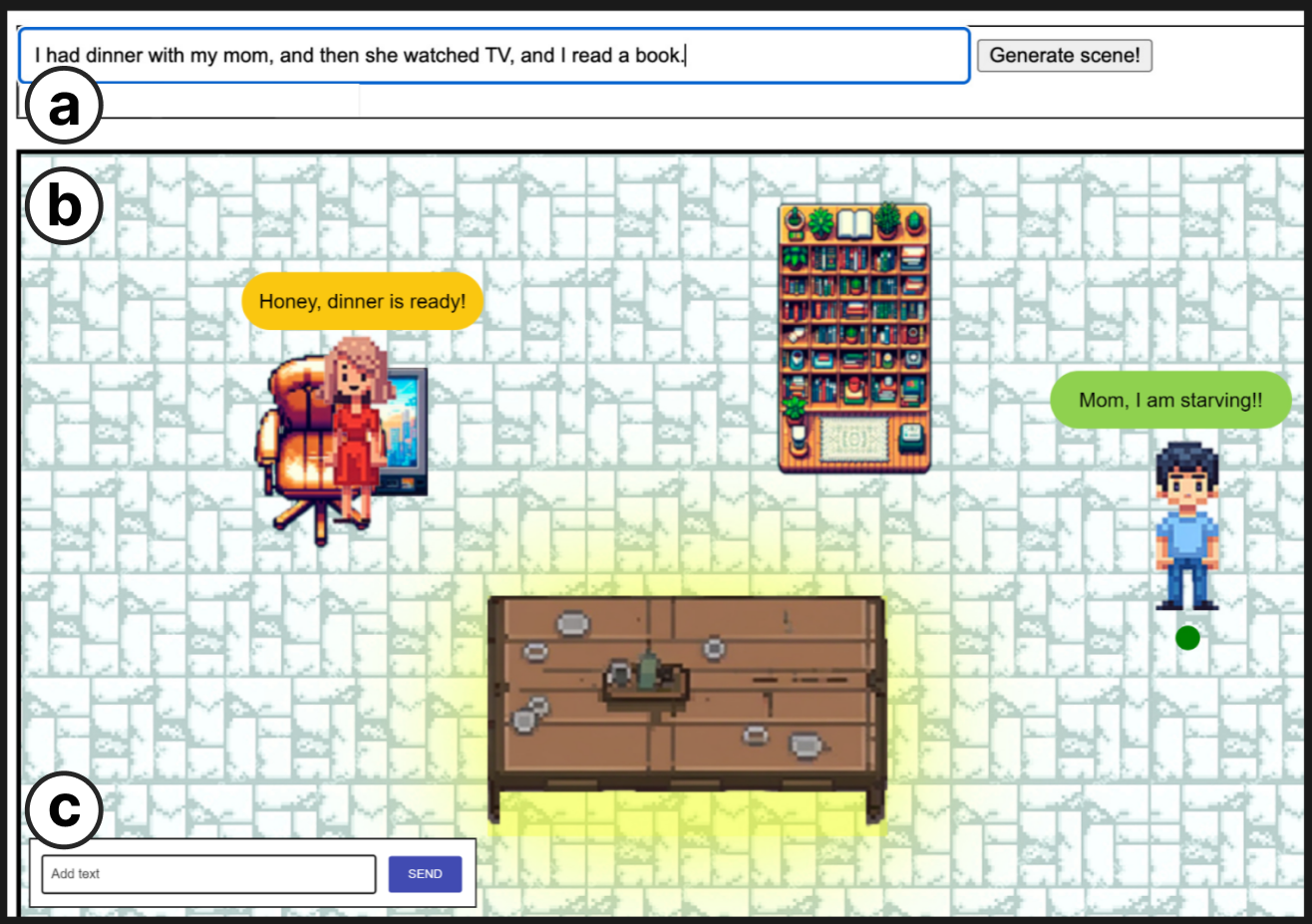}
    \Description{This is the UI of the research probe designed for the formative study.}
    \caption{User interface of the research probe, which includes: (a) a text story input box, (b) a display of the automatically generated scene featuring objects mentioned in the story and characters acting on events, and (c) a chat feature.}
    \label{fig:studytool}
\end{figure}
After a brief introduction, we conducted a semi-structured interview.
The first half of the interview aimed to explore the motivations for authoring and viewing interactive vignettes in everyday storytelling.
We introduced interactive vignettes alongside four common everyday storytelling modalities in social media (single image, sequential images, text, and video), presented in random order with balanced descriptions. 
We asked the participant to rank five modalities from most to least preferred with 1 being best, both as an author and as an audience, allowing us to assess their intrinsic motivation to create and engage with interactive vignettes in everyday storytelling.
From the author's perspective, we inquired about their motivations for authoring and their expected outcomes for an interactive vignette representing everyday stories, assuming they would have adequate authoring supports.

In the second half of the interview, we shifted focus to the participants' expectations for an interactive vignette authoring system. 
The goal of the second half of the formative study was to gather more specific insights about what an ideal output of an interactive vignette authoring system would look like for everyday storytelling, as well as to understand where the authoring system could assist authors in the authoring process.
To ground the discussion in concrete examples, we asked participants to interact with a research probe (Figure~\ref{fig:studytool}) that we developed. 
This probe uses an LLM to automatically convert a short text story into a scene with objects and characters mentioned in the story. 
\rev{
We adopted the three core elements of interactive vignettes (environment, characters, and events) identified in the literature~\cite{Zhao2024}.
Additionally, we drew inspiration from conventional tile-based 2D RPGs (e.g., Pokémon~\cite{pokemon1996}, Chrono Trigger~\cite{chronotrigger1995}) and Generative Agents~\cite{park2023generative}, which demonstrate object-centric character activities and conversational interaction.
Also, we followed the recent trend of storytelling authoring tools (e.g., NarrativePlay~\cite{Zhao2024}, SceneCraft~\cite{kumaran2023scenecraft}) that use natural language story as an intuitive input modality.
In designing the probe, we intentionally kept the affordances minimal yet expressive, so that participants could articulate their expectations, preferences, and the missing features for everyday storytelling.
}

For the analysis of the collected data, the first and second authors performed open coding~\cite{khandkar2009open} on the interview transcripts and observation notes, identifying themes and consolidating them through iterative discussions among the whole research team.
We provide further descriptions of the research probe, study protocol, and interview questions in the Supplementary Material.

\rev{\subsection{Results}}
\rev{We present participants' motivations for authoring interactive vignettes in everyday storytelling, their expectations for the authoring process, and their desired outcomes and viewing experiences.}

\subsubsection{Motivations for Adopting Interactive Vignettes in Everyday Storytelling}
\label{sec:formativeres1}
\textcolor{black}{Participants expressed a desire to both author and view interactive vignettes in everyday storytelling. 
Compared to other modalities, interactive vignettes had a higher average ranking from 
both the authoring ($\mu = 1.77$) and viewing ($\mu = 1.61$) perspectives. 
From the author's side, participants valued the role-playing aspect, which allowed them to \textit{``actively invite the audience to experience my (the author's) perspective and deeply resonate with my everyday story''} (P7).
From the audience's side, participants highlighted the value of \textit{``proactive participation in the events beyond traditional passive consumption of an everyday story''} (P9).
A Friedman's test showed significant differences in modality preference rankings for both authoring ($\chi^{2} = 15.38, p = 0.004$) and viewing ($\chi^{2} = 28.96, p < 0.001$). 
Nemenyi post-hoc tests (Table~\ref{tab:pairwise_authoring} and Table~\ref{tab:pairwise_viewing}) indicate that  from both authoring and viewing perspectives, interactive vignettes ranked significantly better than text (authoring: $\mu = 4.00, p < 0.01$; viewing: $\mu = 4.39, p < 0.01$) and single images (authoring: $\mu = 3.54, p = 0.04$; viewing: $\mu = 3.79, p < 0.01$).
}

\textcolor{black}{Regarding suitable formats for various story types, participants noted that interactive vignette is well suited for everyday stories with clear event sequences (P1, 4, 8, 9)---\textit{``move characters from one event object to another, feels very engaging.''} (P4)---but may be less expressive for stories centered on emotions or inner reflection (P6, 7, 11). Instead, P14 suggested the potential benefits of a text story for such stories, since text stories allow ``putting more details about inner emotions, that are hard to express through visual language and can only be conveyed through words.''}

\begin{table}[h]
\centering
\caption{Pairwise p-values between five modalities were calculated using Nemenyi's test. The cells for significant pairs (\(p < 0.05\)) are bolded. The average ranking for each modality is shown after the modality names. We denote the five modalities as interactive vignette (IV), multiple images (MI), single image (SI), video (VD), and text (TX).}
\vspace{0.2cm}

\begin{subtable}[t]{\linewidth}
\centering
\caption{Authoring side}
\resizebox{\linewidth}{!}{
\begin{tabular}{l|lllll}
\hline
 & \textbf{IV ($\mu$=1.77)} & \textbf{MI ($\mu$=3.08)} & \textbf{SI ($\mu$=3.54)} & \textbf{VD ($\mu$=2.62)} & \textbf{TX ($\mu$=4.00)} \\
\hline
\textbf{IV} & - & 0.22 & \textbf{0.04} & 0.65 & \textbf{< 0.01}\\
\textbf{MI} & - & - & 0.95 & 0.95 & 0.57\\
\textbf{SI} & - & - & - & 0.57 & 0.95\\
\textbf{VD} & - & - & - & - & 0.17\\
\textbf{TX} & - & - & - & - & - \\
\hline
\end{tabular}
}
\label{tab:pairwise_authoring}
\end{subtable}

\vspace{0.4cm}

\begin{subtable}[t]{\linewidth}
\centering
\caption{Viewing side}
\resizebox{\linewidth}{!}{
\begin{tabular}{l|lllll}
\hline
 & \textbf{IV ($\mu$=1.61)} & \textbf{MI ($\mu$=2.96)} & \textbf{SI ($\mu$=3.79)} & \textbf{VD ($\mu$=2.25)} & \textbf{TX ($\mu$=4.39)} \\
\hline
\textbf{IV} & - & 0.15 & \textbf{< 0.01} & 0.82 & \textbf{< 0.01}\\
\textbf{MI} & - & - & 0.64 & 0.75 & 0.12\\
\textbf{SI} & - & - & - & 0.08 & 0.85\\
\textbf{VD} & - & - & - & - & \textbf{< 0.01}\\
\textbf{TX} & - & - & - & - & - \\
\hline
\end{tabular}
}
\label{tab:pairwise_viewing}
\end{subtable}

\label{tab:pairwise_subtables}
\end{table}

\subsubsection{\rev{Expectations for a Lightweight Authoring Process and Automation Support}}
\label{sec:formativeres2}
All participants emphasized they expected the process to remain lightweight. 
As participants self-reported, they usually invested about 20 minutes in creating content for everyday storytelling, setting their expectation for a similar duration in interactive vignette creation. 
Participants considered text-story input intuitive and easy to use.
At the same time, participants perceived the complexity of authoring interactive vignettes and emphasized the need for a \textit{``low-effort''} and \textit{``not time-consuming''} process (e.g., creating various elements of an interactive vignette from scratch is cumbersome), reaffirming the importance of immediacy in everyday storytelling.
As P8 noted, \textit{``Everyday storytelling should be a lightweight task, not too time-consuming or laborious.''}
They therefore appreciated the probe's automation based on natural language text-story input, even though it was not perfect.
As P4 noted, \textit{``changing things is easier than building from scratch''}, which reflects the ``no blank canvas'' concept~\cite{compton2015casual}.

\subsubsection{\rev{Expectations for the Resulting Interactive Vignettes}}
\label{sec:formativeres3}
Participants believed the resulting interactive vignette should convey the main message of their story---who did what, where---consistent with the ``environment, characters, and events'' framework~\cite{Zhao2024} and maintains \textit{``one beginning, one ending''} (P7).
For the \textit{environment}, participants preferred an object-rich design (e.g., \textit{``a sandbox environment like RPGs,''} P9) rather than sparse, randomly positioned objects tied only to specified activities.
For \textit{characters}, participants valued conversational features and expected to represent character personas (e.g., mood, personality, social relationships). 
In contrast, visual details such as clothing and facial features were \textit{``not very important for storytelling''} (P2).
For \textit{events}, participants expected the outcome to reflect the event sequence described in their text-story input and to show characters moving through the scene between objects, which felt \textit{``vivid and immersive''} (P3).
\rev{In sum, participants suggested the authoring system should add object-rich environments with better realism, enable persona building in character construction while retaining the conversational feature, and keep character mobility within the environment for event transitions.}

\subsubsection{\rev{Expectations for Player Divergence in Viewing Experience}}
\label{sec:formativeres4}
Participants emphasized that an engaging viewing experience should involve flexible interaction with objects in the environment and conversations with NPCs. 
When authoring, participants wanted their input story to serve as a \textit{``central skeleton with key events as checkpoints''} (P5), allowing players to go beyond the predefined key events.
As P10 noted, forcing players to strictly follow a fixed author-defined storyline would feel \textit{``too boring, like finishing a homework assignment.''} 
Participants also expected NPCs to respond with adaptive actions when players diverged from the main storyline, rather than remaining idle, to sustain immersion and narrative coherence.

\rev{\subsection{Design Goals}
\label{subsubsec:designgoal}
Based on the findings, we outline three design goals for an interactive vignette authoring system for everyday storytelling.

\noindent\textbf{[DG1] Enable a lightweight authoring process that matches the effort of everyday storytelling practices.}
The system should support single-branch natural-language story input and provide system automation for generating narrative elements.

\noindent\textbf{[DG2] Capture and convey the author's key message in the everyday story.}
The system should enable authors to concretize the core elements (environment, characters, and events) of the interactive vignette and ensure that the generated interactive vignette conveys them.

\noindent\textbf{[DG3] Deliver an engaging viewing experience with situation-aware NPCs in a branching storyline.}
The system should support divergent player interactions and provide adaptive branching narratives.
To preserve the author’s intended events (DG2), the system should use a narrative structure with one beginning and one ending, guiding players through events in the order defined by the author.
}

\section{The \sysname{} System}
\label{sec:system}
\begin{figure*}[h]
    \centering
    \includegraphics[width=1\textwidth]{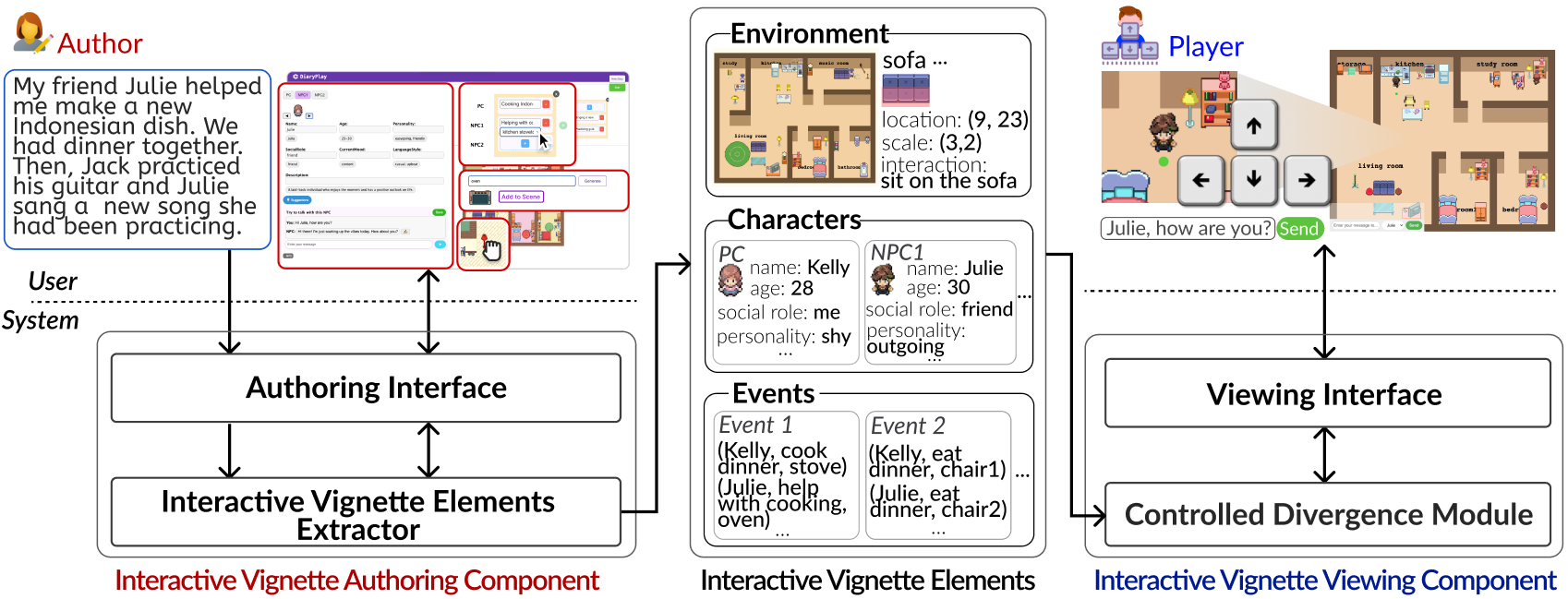}
    \Description{This is the overview of the system.}
    \caption{\sysname{} system overview. The Interactive Vignette Authoring Component takes the author's text story as input, extracts core elements (environment, characters, and events), and guides the author in refining them into structured interactive vignette elements. The Interactive Vignette Viewing Component enlivens NPCs and allows the \rev{player} to control the PC, enabling an interactive narrative experience grounded in the authored storyline.}
    \label{fig:overview}
\end{figure*}

\begin{figure}
    \centering
    \includegraphics[width=\linewidth]{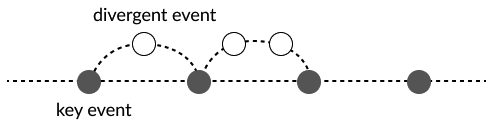}
    \Description{illustration of branch and bottleneck structure.}
    \caption{Illustration of the branch-and-bottleneck structure in interactive storytelling. Solid black circles represent key events that all players encounter, ensuring alignment with the author-intended storyline. Hollow circles represent divergent events, which offer optional or personalized paths that allow for variation in viewing experience. Despite branching paths, the narrative converges at key events.}
    \label{fig:bnb}
\end{figure}

Based on the findings and design goals (Section~\ref{sec:formative}), we present \sysname{}, an AI-assisted interactive vignette authoring system for everyday storytelling. The system comprises two components (Figure~\ref{fig:overview}): (1) the \textit{Interactive Vignette Authoring Component}, which includes the \textit{Authoring Interface} and the \textit{Interactive Vignette Elements Extractor}, and (2) the \textit{Interactive Vignette Viewing Component}, which includes the \textit{Viewing Interface} and the \textit{Controlled Divergence Module}.
To support lightweight authoring (DG1) and ensure the capture of the author's core story message (DG2), the Interactive Vignette Authoring Component guides everyday storytellers in completing and refining interactive vignette elements by surfacing gaps between the natural-language story input and the information required for an interactive vignette. 
To enable adaptive player interactions (DG3) without requiring authors to manually construct branching paths (DG1), the Interactive Vignette Viewing Component automatically delivers branching interactive narratives based on the authored elements. 
We adopt a branch-and-bottle narrative structure (Figure~\ref{fig:bnb}) so that divergent player interactions (DG3) converge back into the author's predefined key events, thereby preserving the author's intended story message (DG2).

\subsection{Interactive Vignette Elements Specification}
\label{subsec:sysspecification}
Based on previous literature~\cite{Zhao2024} and the formative study, we use three core elements to represent interactive vignettes: \textit{environment}, \textit{characters}, and \textit{events}.
To enable lightweight authoring (DG1) while still capturing the main story message (DG2), we designed a minimal but sufficient specification for everyday storytelling regarding each of the three core elements. 
We determined the input fields for each of the elements by referring to previous work and systems, feedback from the formative study, and iterative pilot tests to ensure a reasonably comprehensive representation. 
We note that the purpose of the specification is not to capture comprehensive properties of the three elements that would generalize beyond the everyday story (e.g., a full-scale persona), but to specifically capture the required information to play out the everyday story as a means to minimize authoring effort (DG1).

\begin{itemize}[leftmargin=*,topsep=0pt]
    \item \textbf{\textit{Environment}}: The visual setup of the interactive scene in which the story unfolds. 
    For our system, we use a 25x35 2D grid map with walled spaces (rooms with entrances) to represent a diversity of locations (e.g., office, home, grocery store, theater, etc). 
    The environment includes the layout of the rooms in the scene as well as the placement of the objects.
    Each object in the scene is represented as its location, scale, actions characters can perform with the object, and interaction zones (i.e., where characters have to be spatially when interacting with the object (e.g., on top, in front, next to)).
    \item \textbf{\textit{Characters}}: Persona and visuals of the player character (PC) and the non-player characters (NPCs).
    Based on the character generation forms in existing AI character constructions~\cite{park2023generative, qin2024charactermeet, characterai}, the fields for each of the characters include the name, age, personality, social role (e.g., relationship to other characters, occupation), current mood, language style, examples of conversation snippets for the persona, and character sprites for the visuals.
    \item \textbf{\textit{Events}}: An ordered sequence of \textit{key events}, each of which is a group of simultaneous character activities (e.g., Kelly cooks dinner while Julie helps with cooking, Event 1 of Figure~\ref{fig:overview}). 
    Each character activity is represented as a (\texttt{character}, \texttt{action}, \texttt{object}) tuple, which specifies ``\textit{who} (character) \textit{did what} (action) while \textit{interacting with what} (object)''.
    The order of the key events indicates the chronological order of the events (before/after). We note that not all characters need to have an assigned activity for each key event if not specified in the input story.
\end{itemize}

\vspace{0.03in}
The information available in the specification allows the Interactive Vignette Viewing Component to automatically generate a branch-and-bottleneck narrative structure by enlivening the NPCs who interact with the environment in a way that is consistent with the NPCs' personas and the key events.

\rev{\subsection{Usage Scenario}}
\rev{We illustrate \sysname{} through a usage scenario that walks through both the authoring and viewing scenarios. Kelly, who enjoys sharing her daily moments with friends on social media, uses \sysname{} to create an interactive vignette and share it with her friend Bob, who experiences it as a player.
}

\label{subsec:sysauthoring}

\begin{figure*}
\centering
\begin{subfigure}{0.56\textwidth}
    \includegraphics[width=\textwidth]{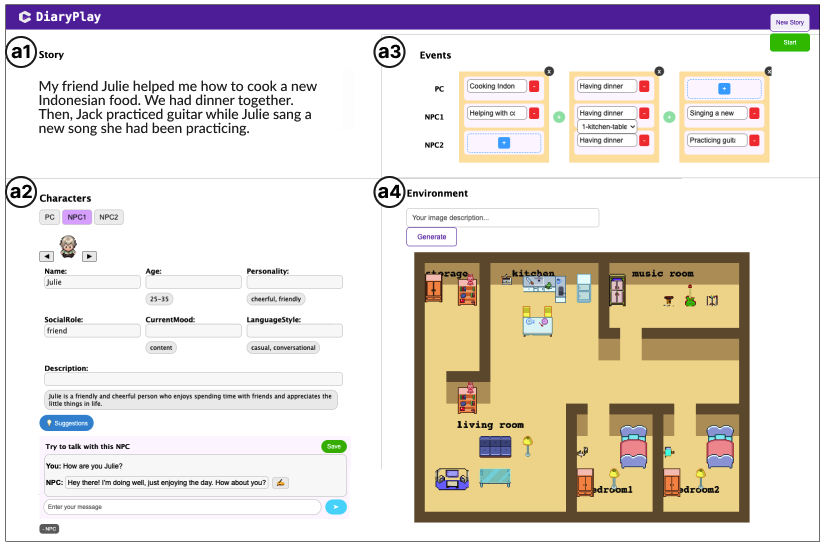}
    \caption{Authoring interface}
    \label{fig:authorUI}
\end{subfigure}
\hfill
\begin{subfigure}{0.43\textwidth}
    \includegraphics[width=\textwidth]{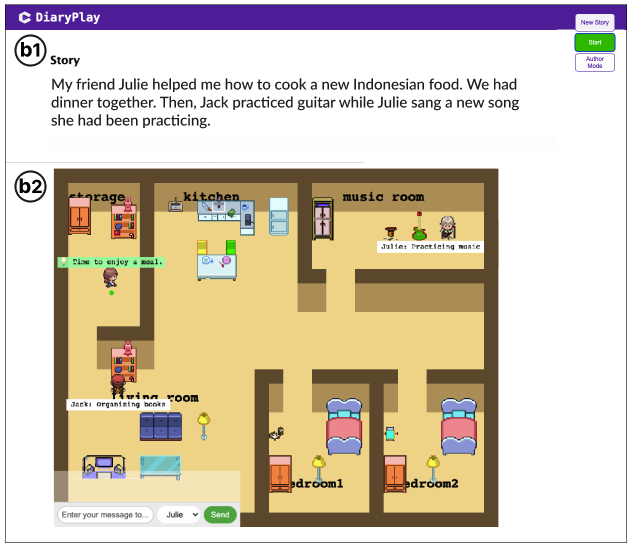}
    \caption{Viewing interface}
    \label{fig:viewerUI}
\end{subfigure}
        
\caption{Authoring interface, including (a1) Text story input, (a2) Characters panel, (a3) Events panel, and (a4) Environment panel; and Viewing interface, including (b1) Text story caption and (b2) Interactive vignette.}
\Description{Authoring interface, including (a1) Text story input, (a2) Characters panel, (a3) Events panel, and (a4) Environment panel; and Viewing interface, including (b1) Text story caption and (b2) Interactive vignette.}
\label{fig:UIs}
\end{figure*}

\subsubsection{\textbf{Authoring Scenario \& Interface}}
\rev{To author an interactive vignette using \sysname{}, Kelly proceeds through the following steps in the \textit{Authoring Interface} (Figure~\ref{fig:authorUI}).}

\begin{figure*}
\centering
\begin{subfigure}[t]{0.47\linewidth}
    \vspace{0pt}
    \includegraphics[width=\textwidth]{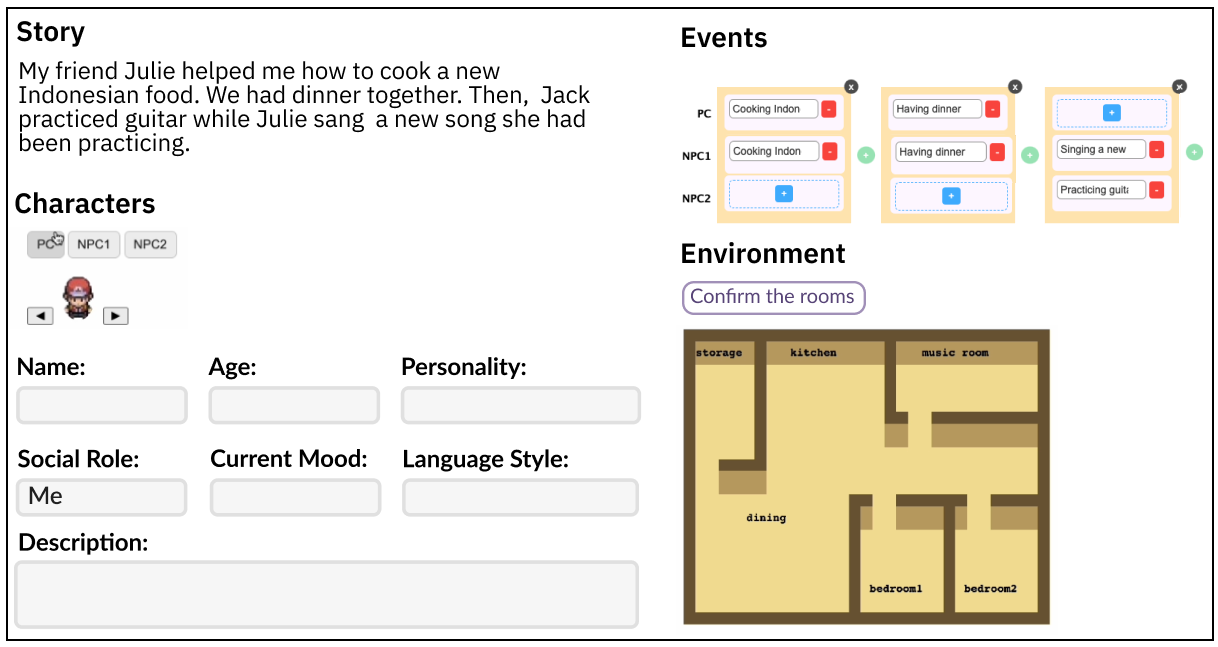}
    \caption{Step 1: Input a story}
    \label{fig:authoringstep1}
\end{subfigure}
\begin{subfigure}[t]{0.25\linewidth}
    \vspace{0pt}
    \includegraphics[width=\textwidth]{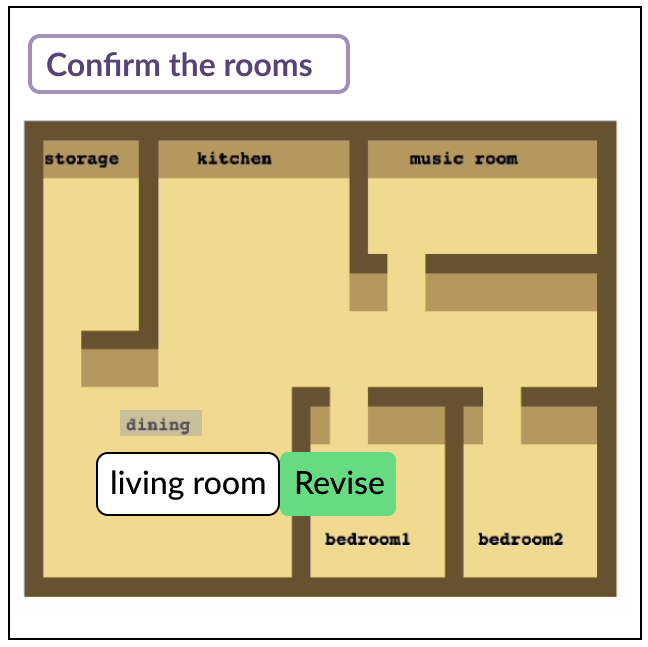}
    \caption{Step 2: Label rooms}
    \label{fig:authoringstep2}
\end{subfigure}
\begin{subfigure}[t]{0.24\linewidth}
    \vspace{0pt}
    \includegraphics[width=\textwidth]{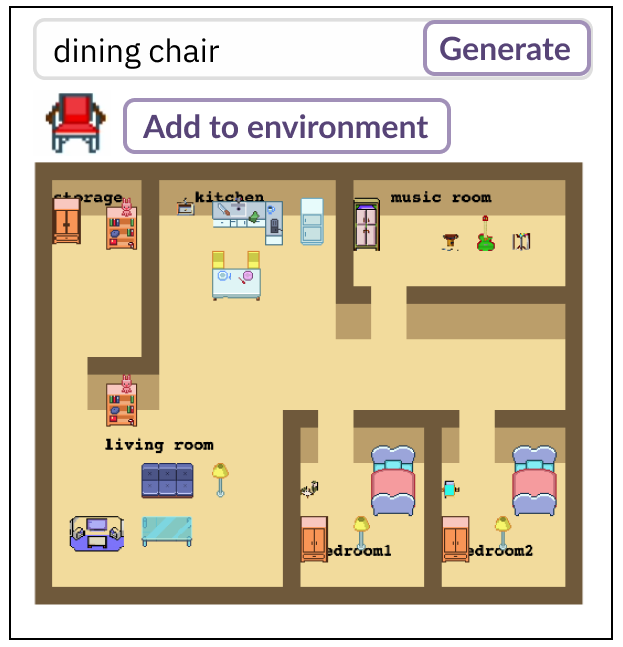}
    \caption{Step 3: Place objects}
    \label{fig:authoringstep3}
\end{subfigure}
\begin{subfigure}[t]{0.48\linewidth}
    \vspace{0pt}
    \includegraphics[width=\textwidth]{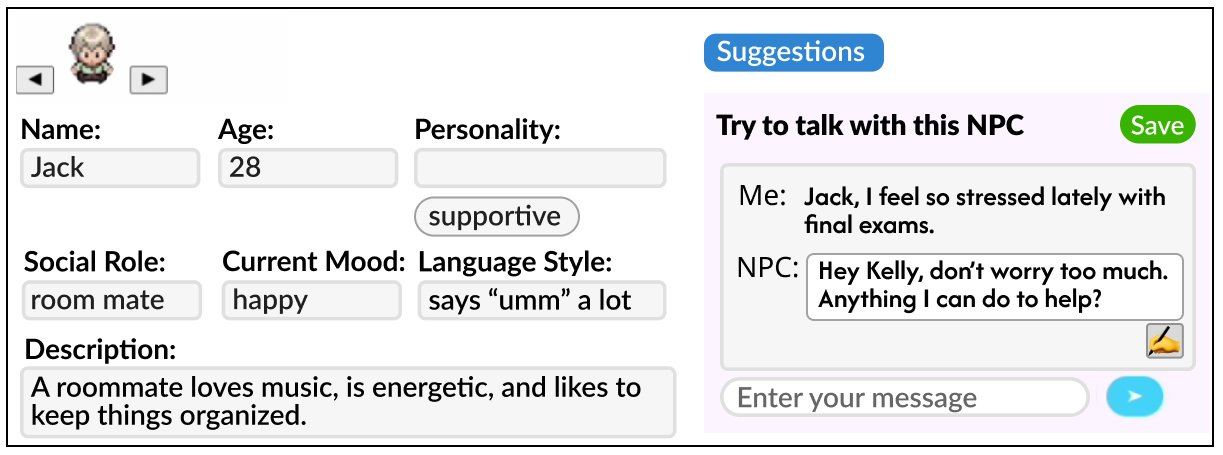}
    \caption{Step 4: Complete the character profiles}
    \label{fig:authoringstep4}
\end{subfigure}
\begin{subfigure}[t]{0.49\linewidth}
    \vspace{0pt}
    \includegraphics[width=\textwidth]{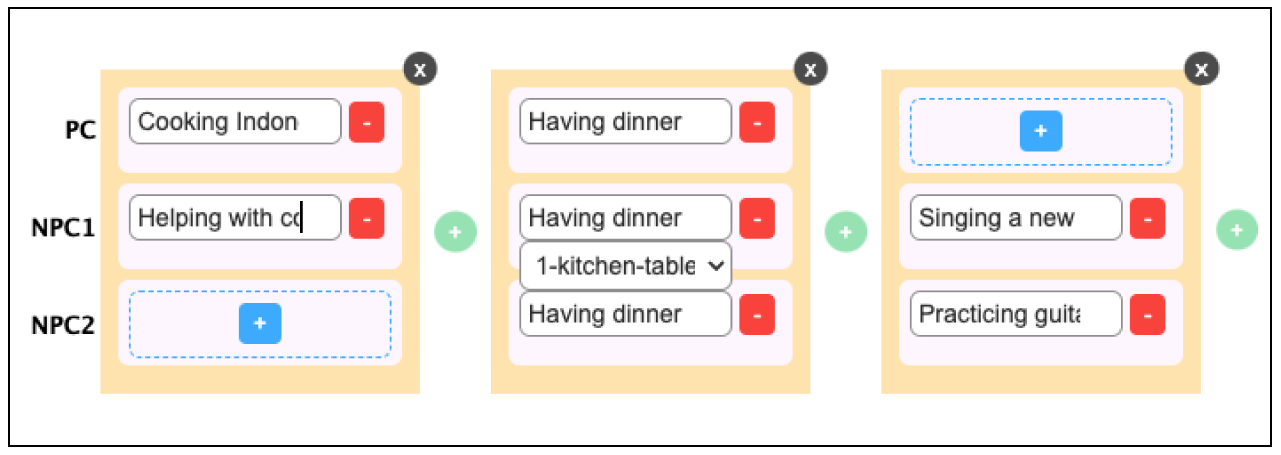}
    \caption{Step 5: Refine the events with object assignment}
    \label{fig:authoringstep5}
\end{subfigure}

\caption{Steps in the Authoring Scenario.}
\Description{Authoring Scenario Steps illustrations.}
\label{fig:authoringsteps}
\end{figure*}

\vspace{0.05in}
\noindent\textbf{Step 1. Input a story.}
Kelly begins by entering an everyday story\footnote{The example is collected from a pilot study.} into the story input box: \textit{My friend Julie helped me make a new Indonesian dish. We had dinner together. Then, Jack practiced his guitar and Julie sang a new song she had been practicing.}
Once the system processes the input, Kelly sees three panels, displaying each of the initial extraction results of three interactive vignette elements (Figure~\ref{fig:authoringstep1}).

\vspace{0.05in}
\noindent\textbf{Step 2. Label rooms.}
Kelly decides to begin with the Environment panel, where she sees a home map layout with several empty pre-labeled rooms (e.g., ``bedroom,'' ``kitchen''; Figure~\ref{fig:authoringstep2}).
She revises the ``dining'' to ``living room'' to better align with her intended setup, then clicks the ``Confirm the rooms'' button to continue.

\vspace{0.05in}
\noindent\textbf{Step 3. Place objects.}
Soon, she sees the rooms automatically populated by the system (Figure~\ref{fig:authoringstep3}), including objects relevant to each room (e.g., a bed for each bedroom) and objects required for the events in the story (e.g., a guitar for Jack to practice).
To customize the environment, Kelly uses mouse drag-and-drop to reposition objects. 
Noticing that the living room feels crowded, she right-clicks to remove a carpet. 
She notices a chair is missing and then decides to add a dining chair in the kitchen, so she types ``dining chair'' in the input box above the map, and clicks the ``Generate'' button, then places the new chair in her desired location.

\vspace{0.05in}
\noindent\textbf{Step 4. Complete the character profiles.}
Next, Kelly moves to the Character Panel. 
She notices three tabs (Figure~\ref{fig:authoringstep4}), each representing a character from her input story: the Player Character (PC), which will be controlled by the player; and two NPCs, her friends Julie and Jack.
Each character tab includes an avatar selection and several blank fields for defining their personas.
Kelly customizes the avatars by cycling through options using arrow buttons.
When filling in the persona fields, she struggles to articulate Jack's personality, so she uses the ``Try to talk with this NPC'' feature to simulate a real-life conversation she had with him.
After reviewing the system-generated response, she directly changed the system-generated dialogue (\raisebox{-\mydepth}{\includegraphics[height=\myheight+4px]{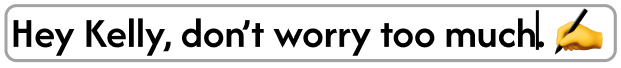}}) to better reflect how Jack would actually speak in the real world.
Once satisfied with the conversation sample, she clicks the ``Suggestions'' button and receives a suggestion bubble beneath the personality field.
Agreeing with the suggestion (``supportive''), she clicks the bubble to automatically fill in the blank field.

\vspace{0.05in}
\noindent\textbf{Step 5. Refine the events with object assignment.}
After completing the Characters panel, Kelly navigates to the Events panel, where she finds a visual timeline of key events (Figure~\ref{fig:authoringstep5}), each containing character activities.
She carefully reviews each event to verify that the system's extracted activities align with her input story.
Kelly noticed that an event generated from ``we had dinner'' included only Julie and her, but it should have included Jack too.
To correct this, she clicks the add button (\raisebox{-\mydepth}{\includegraphics[height=\myheight+1px]{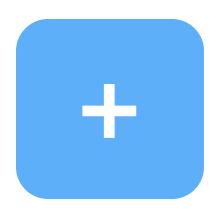}}) and types ``having dinner'' to add it to Jack's activities in the events panel.
After making this revision, Kelly feels that the events are now in-line with her intended story, with character activities arranged chronologically and simultaneous activities grouped within the same event. 
Next, to review and edit the target objects for each activity, she clicks on each activity and sees the system's automatic assignments and opens the drop-down menu to select an object from the list.

Satisfied with the elements of the three panels, Kelly clicks the ``Start'' button to preview her interactive vignette (Figure~\ref{fig:viewerUI}). 
If she wants to refine it further, she can click ``Author mode'' to continue making adjustments.

\subsubsection{\textbf{Viewing Scenario \& Interface}}
\begin{figure*}
\centering
\begin{subfigure}[t]{0.32\linewidth}
    \includegraphics[width=\textwidth]{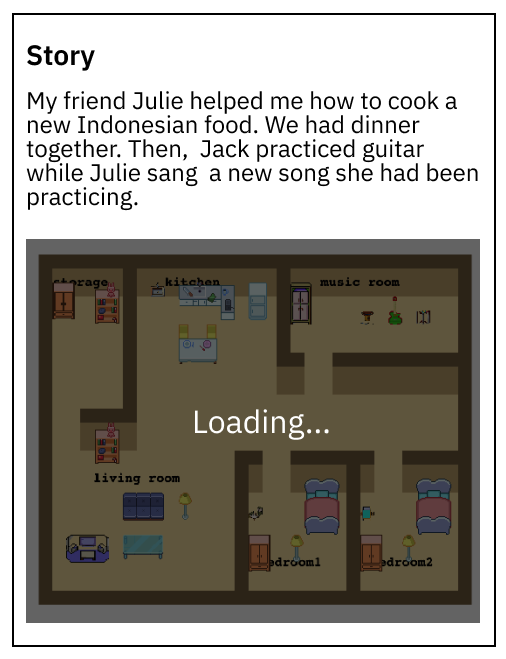}
    \caption{Step 1: Read the caption}
    \label{fig:viewingstep1}
\end{subfigure}
\hfill
\begin{subfigure}[t]{0.31\linewidth}
    \includegraphics[width=\textwidth]{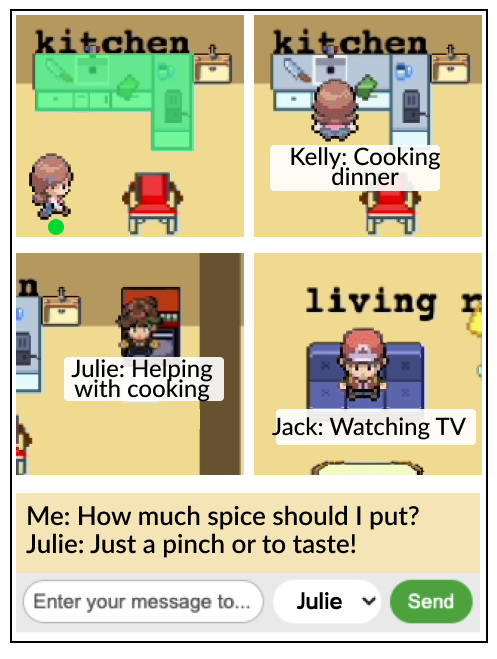}
    \caption{Step 2: Engage with the first key event}
    \label{fig:viewingstep2}
\end{subfigure}
\hfill
\begin{subfigure}[t]{0.31\linewidth}
    \includegraphics[width=\textwidth]{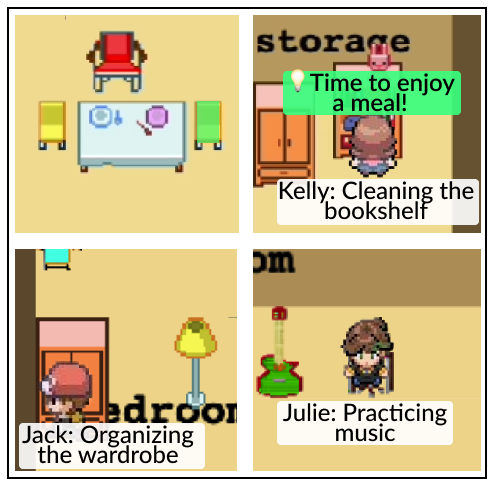}
    \caption{Step 3: Diverge from the key events}
    \label{fig:viewingstep3}
\end{subfigure}
\hfill
\begin{subfigure}[t]{0.32\linewidth}
    \includegraphics[width=\textwidth]{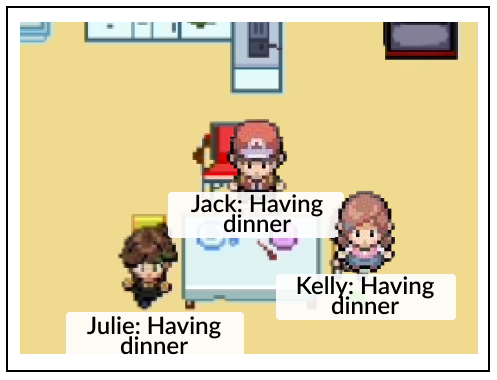}
    \caption{Step 4: Return to the main storyline}
    \label{fig:viewingstep4}
\end{subfigure}
\hspace{4mm}
\begin{subfigure}[t]{0.64\linewidth}
    \includegraphics[width=\textwidth]{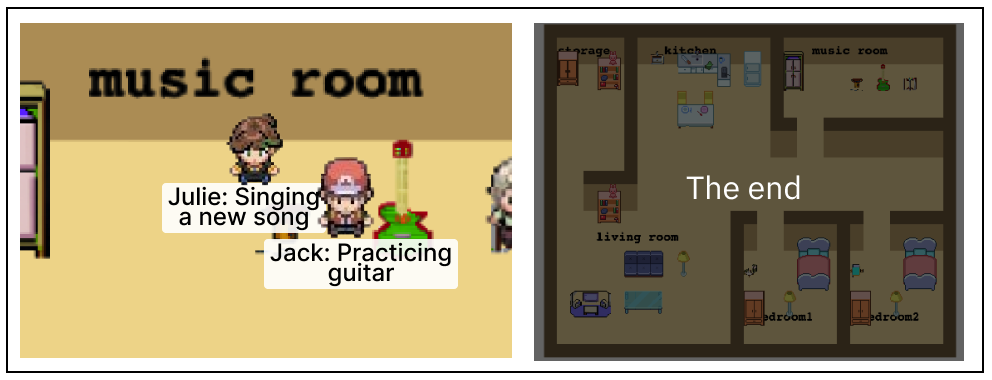}
    \caption{Step 5: Complete the interactive vignette experience}
    \label{fig:viewingstep5}
\end{subfigure}
        
\caption{Steps in the Viewing Scenario.}
\Description{Viewing Scenario Steps illustrations.}
\label{fig:viewingsteps}
\end{figure*}

\rev{After Kelly posts the interactive vignette on social media, Bob plays it through interacting with \textit{Viewing Interface} (Figure~\ref{fig:viewerUI}).}

\vspace{0.05in}
\noindent \textbf{Step 1. Read the caption.} 
Bob starts by reading the text story as a caption (Figure~\ref{fig:viewingstep1}) to get an overview of the characters and events. 
He realizes that he will role-play as Kelly (the PC) and interact with two other characters, Julie and Jack (NPCs).

\vspace{0.05in}
\noindent \textbf{Step 2. Engage with the first key event.} 
As the interactive vignette loads, Bob notices the kitchen stove top glowing (Figure~\ref{fig:viewingstep2}), signaling that an activity is about to take place at the stove. 
To trigger the activity, he uses the arrow keys to move the PC toward it. 
Upon approaching, the activity starts. 
Bob sees the glow disappear, and a speech bubble appears under the PC's avatar to show the PC's activity: ``cooking dinner.''
Bob then observes Julie moving in the kitchen, and her speech bubble reads: ``helping with cooking.'' 
Although the text story does not specify Jack's activity during the cooking, Bob notices that Jack is sitting on the sofa and watching TV rather than remaining idle, giving Bob a sense of character liveliness.
As the cooking event continues, Bob engages with the interactive vignette by role-playing as Kelly and talking with Julie. 
He types into the message box to ask Julie, ``How much spice should I add?'' 
Julie responds, ``Just a pinch or to taste!'' 

\vspace{0.05in}
\noindent \textbf{Step 3. Diverge from the key events.} 
After the cooking event is done, Bob sees the chair glowing, signaling the transition to the next key event --- having dinner. 
However, instead of moving the PC to the glowing chair (Figure~\ref{fig:viewingstep3}) immediately, Bob decides to explore the home environment further. 
He moves the PC to the storage room and triggers the activity ``cleaning the bookshelf'' at the bookshelf.
Interestingly, Bob realizes Jack and Julie are not abandoning the PC to have dinner without Kelly.
Instead, Bob observes that Jack and Julie are doing other activities. 
Jack is ``organizing the wardrobe'', and Julie is ``practicing music.'' 
Meanwhile, Bob receives a subtle inner voice bubble reading: ``Time to enjoy a meal!''
Trying to diverge from eating dinner, Bob types to Jack: ``I want to skip dinner.''
Jack responds, ``Dinner is important. Let's have dinner together,'' guiding the PC toward the next key event.

\vspace{0.05in}
\noindent \textbf{Step 4. Return to the main storyline.} 
Following the inner voice and Jack's response, Bob decides to move the PC to the glowing dining chair.
After the PC arrives at the dining chair, Bob sees Jack and Julie join the PC for dinner (Figure~\ref{fig:viewingstep4}).

\vspace{0.05in}
\noindent \textbf{Step 5. Complete the interactive vignette experience.} 
After the ``having dinner'' event, Bob notices Jack and Julie moving to the music room to practice guitar and singing (Figure~\ref{fig:viewingstep5}). 
With no more glowing objects in the environment, Bob realizes he is free to move the PC, interacting with the characters and objects in the environment. 
After Jack and Julie complete their guitar practice and singing, Bob sees the end screen, marking the conclusion of the experience.

\rev{\subsection{System Implementation}
To support the usage scenario described above, we implement \sysname{} with two modules: (1) the Interactive Vignette Elements Extractor supporting the authoring workflow, and (2) the Controlled Divergence Module running during the viewing experience.}

\subsubsection{\textbf{Interactive Vignette Elements Extractor}}
The Interactive Vignette Elements Extractor takes the author's input story and initially extracts three elements --- environment, characters, and events --- and presents them in the corresponding panels on the Authoring Interface, where the author can review, complete, and modify them with system guidance.

\vspace{0.05in}
\noindent \textbf{Environment Building:}
Systems for interactive storytelling often require authors to construct environments for characters to act within.
To avoid overwhelming authors with the heavy workload of building environments from scratch (DG1), the Interactive Vignette Elements Extractor automatically generates the environment layout and places objects based on the full story input, with each step followed by user confirmation and modification.
\rev{This story-driven environment building echoes Sleep is Death~\cite{sleepisdeath_wiki}, where one player builds the environment for another's story, whereas here the system assists in generating the environment for the author's story.}

To first define the environment layout, the Interactive Vignette Elements Extractor begins by using the LLM to pair the input story with the tags (e.g., residential, retail, office) on one of the layouts (\rev{environment templates with a fixed number of rooms}) designed based on the urban space classification by Pissourios et al.~\cite{pissourios2017classification}.
Then, it uses the LLM to label each of the rooms (e.g., kitchen, bedroom) based on its functionality to support the story input, and asks the author for review and revision.

Once the author confirms the layout and the labels for each of the rooms, the Interactive Vignette Elements Extractor then automatically calculates and populates each of the rooms with \textit{necessary objects} (either event-related (e.g., guitar for Jack's practice) or environment-related (e.g., bed in a bedroom)) using a greedy algorithm. 
If there is still room for additional objects, the extractor generates \textit{decorative objects} that enhance realism (e.g., a lamp in a bedroom). 
The positioning of the objects is based on the LLM's semantic reasoning of object-object relationships (e.g., ``chairs'' close to ``tables''), and object-room relationships (e.g., ``fridges'' next to ``walls''), as well as the path-finding algorithm to ensure the presence of paths to the objects.
Once the population of the rooms is complete, the authors can add, remove, or relocate objects.
\rev{Although there is no fixed limit on room objects, the path-finding algorithm stops additions when the room becomes too crowded for navigable paths.}

For each of the objects, the system assigns possible actions based on the story and appropriate trigger zone type --- where the character sprite needs to be with respect to the object to trigger an interaction --- by prompting the LLM with the object's name to reason how people typically interact with it: \textit{on} (e.g., sleeping on a bed), \textit{partial} (e.g., sitting on the seat portion of a sofa), \textit{around} (e.g., sitting around a table), and \textit{directional} (e.g., opening a fridge from the front side).

\vspace{0.05in}
\noindent \textbf{Characters Construction:}
Based on the interactive vignette elements specification about the characters, the extractor utilizes the LLM to analyze the characters mentioned in the story. 
It identifies \rev{the main character}, first-person narrator (e.g., ``I'') as the PC and classifies all other characters as NPCs. 
For each character, the system extracts explicitly stated persona attributes such as names and social roles, while leaving unspecified traits blank for the author to complete, with system suggestions available to ease authoring. 
Although the LLM is capable of inferring implicit attributes, such as a character's mood, we intentionally avoid auto-filling these to prevent over-guiding the author.

\rev{The conversation-simulation feature aligns with prior work in computational character authoring, which uses LLMs to generate dialogue based on character persona traits~\cite{characterai} and to infer character persona traits based on given dialogues~\cite{qin2024charactermeet}.
The system prompts an LLM to generate NPC replies based on the character's current persona traits, conversation examples, and the story context. 
Authors can edit these responses to better reflect how the real person would speak.
To recommend persona traits, the system prompts an LLM to compare the dialogue against current persona traits: it flags traits contradicted by the conversation, revises them, and proposes new traits implied by the dialogue. 
The recommended list includes retained, revised, and newly inferred traits, which authors can further modify.
By leveraging the interdependence between dialogue and persona traits, \sysname{} supports both explicit and implicit character creation through iterative refinement.
}


\vspace{0.05in}
\noindent \textbf{Events Scheduling:} 
Based on the interactive vignette elements specification about the events, the extractor first prompts the LLM with the characters extracted during character construction and the input story to list the actions performed by each character, matching them to the most appropriate objects in the environment, thereby creating the \texttt{(character, action, object)} tuple. 
Next, the extractor prompts the LLM to group simultaneous activities by multiple characters into a single key event. 
Finally, the system asks the LLM to organize these key events into a time-ordered sequence.


\subsubsection{\textbf{Controlled Divergence Module}}
\label{subsec:sysviewing}
\label{cd}
During the interactive vignette play time, the Controlled Divergence (CD) Module automatically transforms the single-branch sequence of events in the specification into a branch-and-bottleneck narrative structure reacting to the player interactions.
The purpose of the CD Module is to allow both the PC and the NPCs to freely take \textit{divergent activities} (DG3), while being subtly guided to stay within the \textit{controlled boundaries} defined by the storyline from the author (DG2).
From the interactive vignette author's perspective, the CD Module offloads the manual effort of crafting a multi-branch narrative (DG1); from the player's perspective, the CD Module provides player agency within the viewing experience (DG3).

We considered three principles when designing the CD Module.
First, to ensure DG3, the CD Module must provide \textit{real-time} responses to ensure an immersive and coherent narrative experience.
Second, to support DG2, the CD Module should provide \textit{subtle PC guidance} toward the key events when they diverge, to preserve the author-intended storyline.
Third, to support DG2, the CD Module should ensure \textit{believable NPCs} that behave consistently with their personas and the overarching story logic.
\begin{figure*}
    \centering
    \includegraphics[width=0.93\linewidth]{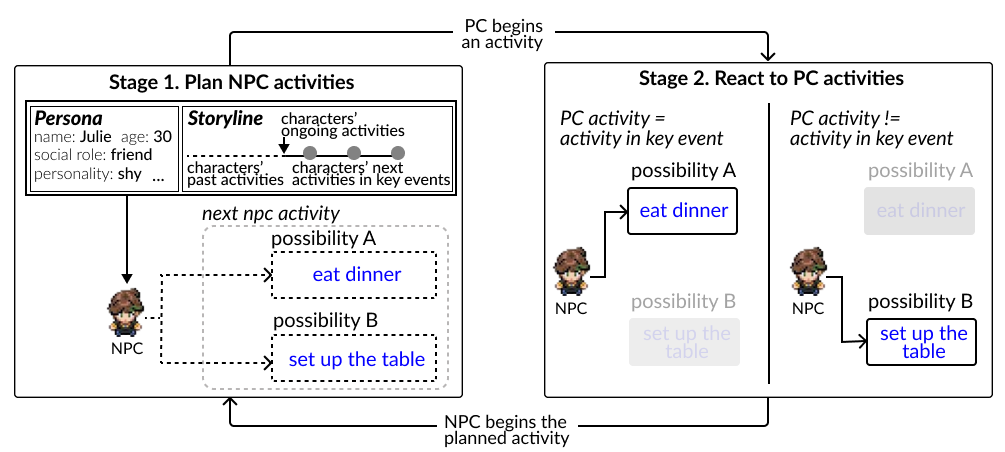}
    \caption{The CD Module plans NPC activities and reacts to PC activities in a two-stage loop.}
    \Description{Illustration of the CD Module plans NPC activities and reacts to PC activities in a two-stage loop.}
    \label{fig:CD}
\end{figure*}
Based on these principles, the CD Module plans NPC activities and reacts to PC activities in a two-stage loop (Figure~\ref{fig:CD}):

\vspace{0.05in}
\noindent \textbf{Stage 1: Plan NPC activities.}
To ensure real-time responsiveness despite the computational delay from activity generation, the CD Module plans NPC activities in advance, triggering planning for the next activity as soon as the current activity begins.
The design of the CD Module takes inspiration from how the human brain copes with the delay between the sensing of the environment and the processing of the information by generating multiple possible action plans and then pruning them according to incoming information~\cite{Kandel2021Visual}.

Specifically, the CD Module generates two \textit{next} activity plans for each NPC using the LLM based on the NPC's \textit{persona} and the \textit{storyline}, which includes all past activities, the characters' ongoing activities and the activities in next key events (Figure~\ref{fig:CD} top): one for when the PC takes action according to the key event (may be defined in the specification; possibility A), and another for when the PC engages in a divergent activity (possibility B).

\vspace{0.05in}
\noindent \textbf{Stage 2: React to PC activities.}
Regarding the NPC activities, the CD Module continuously monitors PC activities and takes the activity A or B based on whether the PC took the activity defined in the next key event (Figure~\ref{fig:CD} bottom).
Once the NPC begins taking the planned action, the CD Module returns to Stage 1 to plan the next activity.

The CD Module also detects the PC's deviation from the key events to guide the player back to the flow of the story through two mechanisms: \textit{inner voice} and \textit{chat-based guidance from the NPCs}.
The inner voice is triggered when the PC does not take the activity defined in the next key event; the CD Module prompts the LLM to write out what the next activity is in a green thought bubble (e.g., Figure~\ref{fig:viewingstep3}; ``Time to enjoy a meal'').
When the PC's chat shows an intention to deviate too far from the key events, the CD Module uses the NPCs' chats to guide the PC back.
For example, if the PC attempts to skip a key event (e.g., ``Can I skip dinner?''), the NPC would not agree to the PC but rather provide a response suggesting returning to the author-intended storyline, ``Dinner is important. Let's enjoy it together.''
\rev{
Instead of forcing players back to the key event, the system uses inner-voice cues and dialogues to nudge players~\cite{schneider2017nudging}. 
To preserve immersion, 
we 
allow divergent activities to continue as long as needed without a fixed limit.
}

\subsection{Implementation Notes}
We developed \sysname{} with Phaser~\cite{phaser}.
The visual assets are sourced from online open resources~\cite{itchio} and generated by DALL·E 2~\cite{dalle2}. 
We used OpenAI's gpt-4o-2024-11-20~\cite{gpt4o}. 
We include LLM prompts in the Supplementary Material. 
To manage computational costs, the current version of \sysname{} supports a maximum of three characters. 
To prevent harmful conversations, the CD Module monitors dialogues and withholds responses if they violate OpenAI's usage policies.

\section{Preliminary Technical Evaluation}
\label{ter}
The NPC activities generated by the CD Module are central to delivering an immersive and interactive experience for players. 
In this preliminary technical evaluation, we focused on the CD Module's ability to generate believable NPC activities, since believability is widely recognized as a crucial aspect of narrative quality~\cite{gomes2013metrics, curtis2022toward, lee2012you}.
We performed an IRB-approved preliminary technical evaluation with five conditions:
\begin{itemize}[leftmargin=*,topsep=0pt]
    \item \textbf{CD Module Condition [CD]}: We use the CD Module of our system to generate NPC activities based on persona and storyline properties.
    \item \textbf{Baseline Condition [BL]}: We use a simple baseline module that randomly selects an object in the environment and prompts the LLM to assign an NPC activity. 
    \item \textbf{Human Author Condition [HA]}: We asked the original author of each of the interactive vignettes used in the study to directly fill in the NPC activities.
    This condition serves as an optimal solution to the activity generation problem.
    \item \textbf{Persona Only Condition [PO]}: We generate NPC activities with our CD Module with only the persona and without the storyline property. 
    \item \textbf{Storyline Only Condition [SO]}: We generate NPC activities with our CD Module with only the storyline and without the persona property. 
\end{itemize}  
With these conditions, we consider two hypotheses: \\
\textbf{[H1]}: The CD Module is capable of generating NPC activities more believable than the baseline and on par with the human authors themselves. \\
\textbf{[H2]}: Both the persona and the storyline contribute to generating more believable NPC activities.

We note that the preliminary technical evaluation was performed with an earlier version of the system that only included limited environment editing features.
However, the system included complete support for character and event extraction, which are highly relevant for the preliminary technical evaluation.

\subsection{Method}
To first obtain the interactive vignettes that we could generate the NPC activities on for each of the five conditions, we recruited 16 participants (7 females and 9 males, aged 19 to 30) within KAIST through recruitment posts.
Each participant authored an interactive vignette with \sysname{}. 
Next, another participant interacted with the interactive vignette as the player; everyone triggered the CD Module to either fill in unspecified activities or divergent activities (on average three times).
We then showed the character-activity table (Table~\ref{tab:techeval}) to the original author of each interactive vignette with the CD Module's generated NPC activities redacted and collected the author's intended NPC activities for the HA Condition.
For the two ablation conditions (PO, SO), we used the appropriate variants of the CD Module to generate NPC activities.

Next, to compare the conditions, we recruited a disjoint set of 16 evaluators (6 females and 10 males, aged 18 to 33) within KAIST through recruitment posts.
After introducing the character-activity table, we provided evaluators with the character attributes and input story as supporting material, along with the character-activity tables generated under five different conditions. 
We then asked them to rank the believability of the NPC activities (highlighted in green in Table~\ref{tab:techeval}) across these five conditions (1 is the best, 5 is the worst).
Note that evaluators saw activity tables generated in all 5 conditions in random order without notations.
We then asked the evaluators to leave any free-form comments on each of the rankings.
We did not let technical evaluators experience interactive vignettes to minimize confounding distractions (e.g., visuals) beyond text-based NPC activities.
Throughout the evaluation, we did not describe how the activities were generated and randomized the order in which the five conditions were shown to avoid potential biases.
Each evaluator ranked the NPC activities from 12 sets of conditions; we collected a total of 192 sets of rankings.

Each of the participants and the evaluators received 25000 KRW ($\approx 19$ USD) for 2 hours of their time. 
We include detailed preliminary technical evaluation materials in the Supplementary Material. 

\subsection{Results}
\begin{table}[t]
\centering
\caption{A character/activity table prints the activities of all characters in a temporal order. NPC activities generated during the interactive vignette viewing period, or added later by human authors, are highlighted in green.}
\resizebox{\linewidth}{!}{
\begin{tabular}{|l|p{2.5cm}|p{2.5cm}|p{2.5cm}|}
\hline
Events & \textbf{PC (Me)} & \textbf{NPC1 (Name)} & \textbf{NPC2 (Name)} \\ \hline
1 & 
(sleep, bed) &
(cook dinner, oven) &
\cellcolor[HTML]{D9EAD3}(watch TV, sofa) \\ \hline
2 & 
(eat dinner, chair1) &
\cellcolor[HTML]{D9EAD3}(eat dinner, chair2) &
\cellcolor[HTML]{D9EAD3}(eat dinner, chair3) \\ \hline
... & ... & ... & ... \\ \hline
n & 
(shower, bathroom) &
(painting, art desk) &
(do laundry, dryer) \\ \hline
\end{tabular}
}
\label{tab:techeval}
\end{table}

\begin{table}[h]
\centering
\caption{Pairwise p-values between five conditions were calculated using Nemenyi's test. The cells for significant pairs (\(p < 0.05\)) are bolded. The average ranking for each condition is shown after the condition names.}
\resizebox{\linewidth}{!}{
\begin{tabular}{l|lllll}
\hline
 & \textbf{CD ($\mu$=2.37)} & \textbf{BL ($\mu$=4.00)} & \textbf{HA ($\mu$=2.58)} & \textbf{PO ($\mu$=2.74)} & \textbf{SO ($\mu$=3.31)} \\
\hline
\textbf{CD} & - & \textbf{< 0.01} & 0.71 & 0.14 & \textbf{< 0.01}\\
\textbf{BL} & - & - & \textbf{< 0.01}& \textbf{< 0.01}& \textbf{< 0.01}\\
\textbf{HA} & - & - & - & 0.84 & \textbf{< 0.01}\\
\textbf{PO} & - & - & - & - & \textbf{< 0.01}\\
\textbf{SO} & - & - & - & - & - \\
\hline
\end{tabular}
}
\label{tab:pairwise}
\end{table}

The average rankings of the character activities generated for each condition were 2.37 for CD, 2.58 for HA, 2.74 for PO, 3.31 for SO, and 4.00 for BL.

A Friedman's test shows that there exist significant differences in the ranking of the believability of the conditions ($\chi^{2} = 133.78, p < 0.001$).
To understand where the pairwise differences are, we apply the Nemenyi post-hoc test (Table~\ref{tab:pairwise}); we analyze each of the differences further along each of the hypotheses.

\subsubsection{Assessing H1.}
The character activities generated by the CD Module ($\mu = 2.37$) ranked significantly better than those generated in the baseline condition ($\mu = 4.00$; $p < 0.01$). 
As E3 commented, character activities generated by the CD Module were \textit{``more temporally logical and relevant to previous and later activities''} and \textit{``captured established characters well,''} but those generated in the baseline conditions were \textit{``quite random, did not make much sense.''} 

We did not find a significant difference between the ranking of character activities generated by the CD Module ($\mu = 2.37$) and human authors ($\mu = 2.58$; $p = 0.71$). 
This result indicates that the CD Module generates character activities on par with human authors regarding the believability.
In sum, the results support [H1]: the CD Module generated more believable NPC activities than the baseline and was on par with the human authors. 



\subsubsection{Assessing H2.}
The character activities generated in the PO condition ($\mu = 2.74$) ranked significantly better than those generated in the baseline condition ($\mu = 4.00$; $p < 0.01$). 
This result suggests that persona contributes to believable character activity generation.
In addition, the character activities generated in the SO condition ($\mu = 3.31$) ranked significantly better than those generated in the baseline condition ($\mu = 4.00$; $p < 0.01$). 
This result suggests that the storyline contributes to believable character activity generation.

The character activities generated by the CD Module ($\mu = 2.37$) ranked significantly better than those generated in the SO condition ($\mu = 3.31$; $p < 0.01$), suggesting that combining persona with storyline leads to more believable activity generation than using storyline alone.
The character activities generated by the CD Module ($\mu = 2.37$) ranked better than those generated in the PO condition ($\mu = 2.74$; $p = 0.14$) on average, although we did not see a significant difference between PO and CD.
In sum, the results support [H2]: both the persona and the storyline contributed to generating more believable NPC activities.

\section{User Study}
\begin{figure*}
    \centering
    \includegraphics[width=1\textwidth]{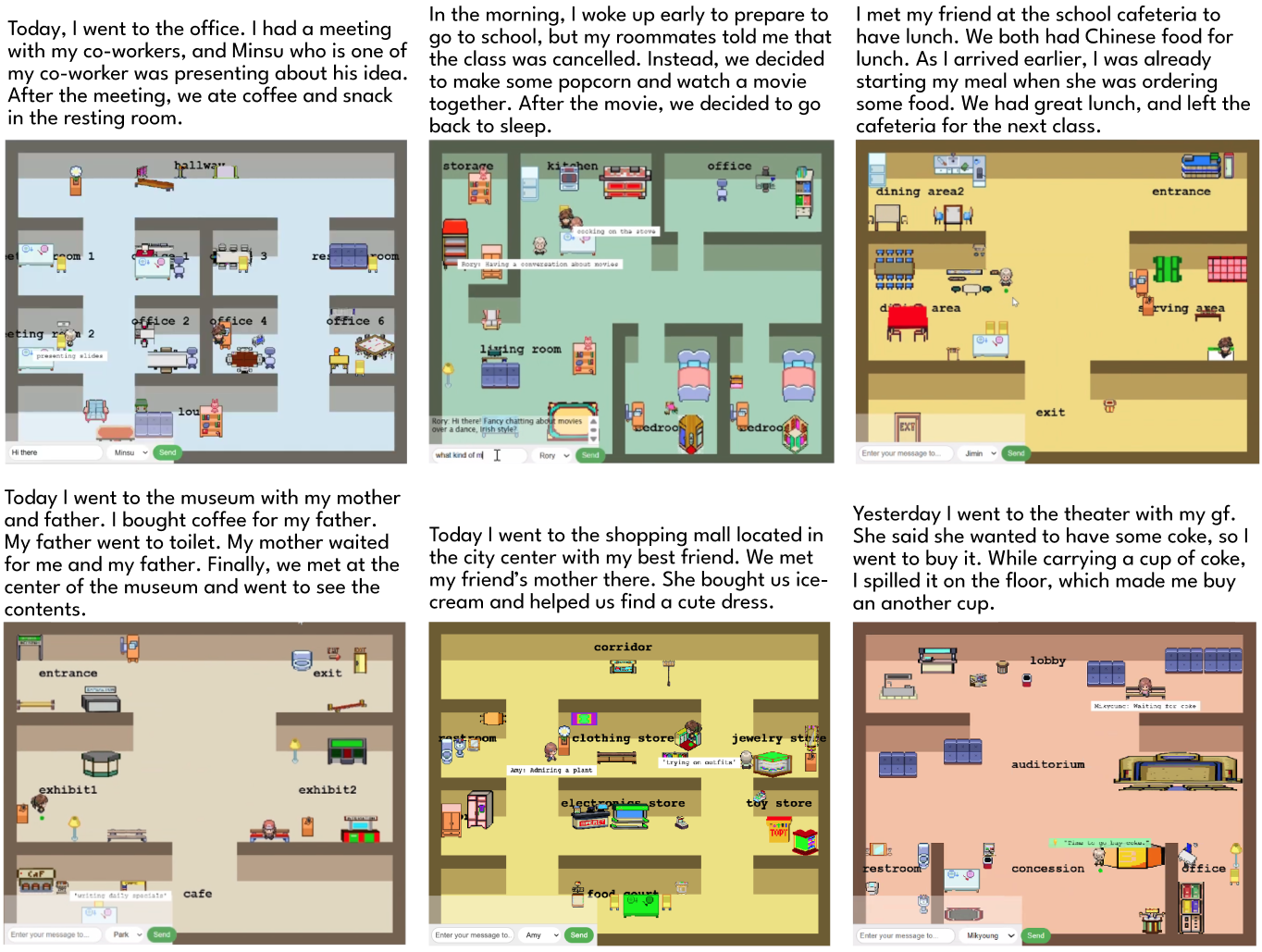}
    \caption{Example interactive vignettes created by participants from the user study. Minor grammatical errors in the original stories have been revised for illustration.}
    \Description{Some interactive vignettes authoring results from the user study.}
    \label{fig:result}
\end{figure*}
\label{sec:userstudy}
We conducted an IRB-approved user study to understand the authoring and viewing experiences of interactive vignettes using \sysname{} for everyday storytelling.
Specifically, we gathered qualitative feedback about the system around the research questions (RQs): \\
\textbf{[RQ1]}: How well does \sysname{} assist interactive vignette authoring for everyday storytelling?\\
\textbf{[RQ2]}: How well does \sysname{} capture and convey the author's main story messages?\\
\textbf{[RQ3]}: How is the viewing experience delivered by the interactive vignette generated with \sysname{}?\\

\subsection{Participants}
We recruited 16 participants (3 females and 13 males, aged 18 to 26) through recruitment posts within KAIST
(\rev{See detailed participants' background information in Table~\ref{tab:userstudyparticipants} at Appendix~\ref{app:participant}}).
\rev{Participants possessed various academic backgrounds,
and were active social media users, creating and consuming everyday storytelling content (e.g., image posts, journaling, text updates, personal vlogs) more than once per week. 
Many participants reported experience with AI-assisted authoring tools for text and image generation.}

In each session, we randomly paired two participants and had them alternate roles as author and player: the player shared their takeaway messages, and the author responded on whether they matched the core story message (who did what, where),~\rev{an approach used in evaluating authoring tools~\cite{louie2022expressive}}.
Each user study session lasted around 1.5 hours, and each participant received 30000 KRW ($\approx 21$ USD) as compensation for their time.

\subsection{Procedure}
We began each in-person user study session by introducing the study's purpose, key terms, and a tutorial on using \sysname{} (about 5 minutes). 

Following the introduction, we let each participant create an interactive vignette based on their personal daily moment, single-location story with \sysname{} on the provided device independently.
Throughout the authoring process, we asked the participant to think aloud while interacting with the \sysname{} authoring interface. 
With participant consent, we recorded the on-screen actions and took observation notes of their authoring behaviors.
\rev{We scheduled up to 30 minutes for the authoring task and informed participants that they could work at their own pace within this time frame. All participants completed the task within 20 minutes.}

After the participant completed the authoring task, we conducted a semi-structured interview (around 20 minutes) about their authoring experience.
We asked about their overall experience, perceived workload, the effectiveness of system features in authoring assistance, satisfaction with the created interactive vignette, and suggestions for system improvements in supporting interactive vignette creation for everyday storytelling.

Then, we invited the two participants to experience each other's interactive vignette as players.
\rev{We informed participants that they need to play the interactive vignette from beginning to end, but were free to replay it as many times as they wished. 
We then ended the viewing task to ensure that the final half hour remained available for the remaining tasks.}
We asked the player to report their takeaway messages of the characters, events, environment, and feelings after the viewing experience, and we asked the author to comment on how well it aligns with their intended narrative. 

Finally, we conducted a semi-structured interview (around 20 minutes) about participants' viewing experience. 
We asked what they liked and disliked about the interactive vignette delivered by \sysname{} and invited suggestions for improving how the system presents interactive vignettes in everyday storytelling.

To analyze the data, we used open coding~\cite{khandkar2009open} to inductively analyze participants' behaviors and reflections, with particular attention to how they related to DGs but open to unexpected insights.
The first, second, and third authors reviewed the entire dataset, which included (1) interview transcripts, (2) observational notes, and (3) screen recordings. 
Each author then independently coded the data using both \textit{in vivo} and \textit{pattern coding} approaches. 
After the initial coding, we organized the codes into preliminary themes by merging our interpretations. 
Through ongoing team discussions, we refined the themes, aligning them with our research questions.

\subsection{Results}
\label{subsec:results}
Overall, our findings show that \sysname{} enables authors to create interactive vignettes that express their intended everyday stories, while balancing an engaging viewing experience with manageable authoring effort.
As illustrated in Figure~\ref{fig:result}, participants used \sysname{} to express a wide variety of stories, spanning across various locations (e.g., office, library, museum, home, theater, shopping mall), characters (e.g., family members, friends, colleagues, strangers, classmates), and types of events (e.g., work day, school morning routine, museum visit). 
This demonstrates the \sysname{}'s capability to support diverse expressions of everyday stories.

\subsubsection{Answering RQ1.}
Overall, participants perceived the duration and workload of creating an interactive vignette as appropriate for everyday storytelling, indicating that the system fulfilled DG1.
Specifically, participants completed interactive vignettes in 12–19 minutes ($\mu$ = 16) with \sysname{}, a duration they regarded as reasonable for everyday storytelling.
They appreciated the system's automatic and mostly accurate transformation from text story to interactive vignette elements: it eased the challenge of starting from scratch and provided \textit{``a great foundation for further authoring''} (P1).

However, the system's parsing of input stories did not always align with participants' intentions. 
Two participants encountered errors in how the system parsed events, stemming from ambiguity in the story inputs. 
P13's story input had ``we ate chicken together,'' which the system parsed as two people having a meal, though he had intended three. 
P9's phrases ``had a meeting'' and ``talked about research topics'' were meant as two separate events, but were initially parsed as one. 
Despite such errors, both participants felt that corrections were easy and fast to make in the authoring interface.
Notably, P13 initially attempted to resolve the issue by editing the input text itself---rewrote the story as ``three of us ate chicken together'' and expected the event element to update automatically while preserving existing content. 
This behavior points to the potential of a more fluid authoring workflow, where modifications to the input story seamlessly propagate into the interactive vignette.

Beyond error correction in events, \sysname{} also supported the participants to refine the generated environment and characters to better express their intentions. 
In environment design, participants repositioned three to five objects to improve realism or visual appeal. 
For instance, P11 moved a sofa closer to the TV and aligned two chairs to make the space feel more \textit{``realistic and visually appealing''}. 
P14 noted: \textit{``I felt I was doing something creative while arranging the environment, but it wasn't burdensome. I didn't need to draw assets---just by the text input, moving, adding, or deleting objects, I saw my imagined space appear in reality.''}
The existing modification features satisfied most participants, while a few participants (3/16) desired finer control over objects, such as rotation and resizing.

We observed that a substantial portion of authoring time, about half on average, was devoted to character building.
P10 explained: \textit{``Text input alone is too weak to express characters, especially their mood and personality. Building the persona gave me a chance to develop them more deeply.''}
Additionally, participants found the process enjoyable, with P11 describing it as the \textit{``most fun and creative step.''}
When filling in persona attributes, many participants (9/16) found it difficult to articulate personality traits without guidance. 
Thanks to the \sysname{}'s conversation simulations and trait suggestions, they felt it was getting easier to build the persona. 
Other than being directly used by the system as examples when generating NPC dialogues with the player, the conversation simulations helped authors obtain a clearer understanding of how the character was understood by \sysname{}, and they can make further modifications to the persona. 
Moreover, the persona trait suggestions based on the conversation simulations helped authors fill in the persona form.
P4, for example, initially struggled to \textit{``pick the right words to describe a friend's personality beyond bland `good' or `bad'''}, but through conversation simulations, the system suggested the trait ``supportive,'' which he found to be highly relevant and used it. 
Beyond the current system design, some participants (4/16) expressed a desire for more nuanced emotional expression. 
Currently, authors can assign only a static ``current mood'' to a character, but participants wanted support for modeling dynamic emotional progression. 
As P5 put it: \textit{``I wish I could design how a character's emotions change in the story. For example, happy at first, then sad, then happy again.''} and suggested a feature for dynamic emotional modeling that evolves in sync with story events.


\subsubsection{Answering RQ2.}
Through reviewing their own interactive vignettes and hearing takeaways from paired players, participants felt that \sysname{} successfully captured and conveyed their core story message by reflecting character personas and the intended sequence of events within the designed environment, thereby satisfying DG2.
After completing the authoring process, participants previewed their interactive vignettes and confirmed that the outcome preserved their intended message, accurately reflecting key events and character persona. 
For example, P14 remarked, \textit{``I'm satisfied that my character speaks the way I defined. I specified in the persona that she often says `oh really,' and now her dialogue reflects that perfectly.''}
Participants also felt their intentions were preserved when players shared their takeaways, even though each player engaged in at least one divergent activity. 
For instance, P7 viewed P8's interactive vignette about three schoolmates meeting for a project in the library. 
Instead of joining the meeting immediately, P7 had the main character read a book from a shelf, while observing the two NPCs ``turning on the monitor'' and ``preparing slides'' in the meeting room. 
He then joined the meeting and talked with the NPCs, still following the core storyline. 
As P8 reflected, \textit{``player understood what happened in my story, though there were divergent interactions. I think they are tiny but nice decorations. And the player also learned well about the characters, like their speaking tones and personality, through chatting with the NPCs.''}

At the same time, authors expressed a desire to review a play log of how players interacted with their interactive vignettes. P8 noted: \textit{``I'm quite curious what exactly happened on the player's end. I felt that as the author, I should be aware of what my audience experienced, even if the core events remained the same.''}
Envisioning real-world use where their interactive vignettes might be viewed by a broader audience, P15 similarly expressed a desire to collect and review all viewing experiences, describing this as an \textit{``author–player interaction, similar to social media posts.''}

Participants also found the original text story helpful for understanding the author-defined main storyline. 
However, a few participants (P3, 5, 10) noted mismatches between the text story and what was actually unfolding in the interactive vignette. 
This discrepancy arose because the text story included the key events, omitting the divergent paths introduced during interaction. 
As P10 remarked, \textit{``I felt the text story was somewhat of an overview, but not a real caption.''} 
At the same time, participants expressed the need to see a transparent distinction between human-authored and system-generated content. 
As P10 noted, \textit{``Though NPC's activities and their dialogues were guided by the human author's intent, they were system-generated, not human-generated directly. As a player, I would like to clearly know that. It should note which parts are system-generated and which are human-made, to avoid misunderstanding.''}
While participants generally appreciated the human-AI co-creative nature of the interactive vignettes, they felt that a lack of transparent authorship designation could lead to confusion.

\subsubsection{Answering RQ3.} 
All participants described interactive vignettes as engaging and valued them as a playful, memorable medium for everyday storytelling. They especially enjoyed the liveliness of NPCs and the chance to explore alternative branches within the same story, demonstrating that the system successfully met DG3.
Participants appreciated the novelty of stepping into the author's perspective and actively exploring the everyday experience through unfolding events. 
Most participants (14/16) described this format as more immersive and emotionally resonant than typical social media content such as text, images, or videos.
As P15 explained, \textit{``On social media, I usually just browse others' daily updates quickly, but viewing the interactive vignette, I felt I was actually experiencing their daily moments, which let me resonate with it a lot.''} 

Participants especially appreciated the liveliness of NPCs, who remained active and responsive even during player-initiated divergent activities. 
P11 noted, \textit{``In RPGs I've played, some NPCs just stand there and feel oddly lifeless, but in my viewing experience, I felt like all the characters were thinking and alive.''}
Many participants (12/16) revisited the interactive vignette multiple times to explore alternative interactions between key events. 
As P7 noted: \textit{``It's fun to explore these little deviations, like opening up a few multiverses while still sharing the same core storyline.''}
Similarly, P14 tried different conversations across playthroughs and described the experience as \textit{``feeling like living inside the story.''}

\section{Discussion}
We discuss (1) the balance of human authoring and system generation in everyday storytelling, (2) opportunities and considerations for integrating DiaryPlay into social media, and (3) ethical considerations for AI-generated content in everyday storytelling. 


\subsection{Balancing Human Authoring and System Generation in Everyday Storytelling}
From the design and evaluation of \sysname{}, we derive several key design lessons.
First, everyday storytellers should be able to modify intermediate results generated by the system.
Our user study findings confirmed that the everyday storytellers found it useful to see and modify the system-generated narrative elements as intermediate results.
Initially, we designed the interactive vignette specification dashboard as a way to surface the information gap between the text story input and what is required to construct an interactive vignette (e.g., missing character personas or unspecified event objects).
In practice, user study participants used the specification not only to fill in missing information, but also to verify whether the system's interpretation aligned with their intent and to correct misrepresentations.
This aligns with prior findings on human–AI cocreation~\cite{shi2023understanding}, which highlight the importance of exposing AI-generated intermediate results to support iterative refinement and to enhance human authors' visibility, flexibility, and sense of agency.

At the same time, everyday storytellers should be able to direct the authoring of interactive vignettes without needing to specify every narrative detail.
User study participants appreciated being able to define key events and characters, and let the system automatically generate adaptive NPC activities and dialogues at interactive vignette runtime.
This authoring approach enabled everyday storytellers to support player divergence within an intended narrative boundary (DG2, DG3), without requiring them to manually craft every narrative branch (DG1).
Traditionally, supporting divergent player interactions has required extensive multi-branch authoring~\cite{lu2025whatelse,linaza2004authoring,calderwood2022spinning,leandro2024geneva}, which emphasizes authors' full control over every narrative branch and is incompatible with lightweight everyday storytelling. 
In contrast, the CD Module enabled a new human–AI co-authoring approach that balances author agency with low authoring effort: authors define the narrative boundary, and the system generates adaptive branches within it.
This approach can extend beyond NPC behavior generation in interactive vignettes to other forms of everyday storytelling, such as generating text-based branches for interactive text stories~\cite{twine} or image-based story material for interactive collages~\cite{downpourTool}.

Importantly, the balance between the human authoring and system generation---as demonstrated in \sysname{}---was intentionally tailored for the lightweight, informal nature of everyday storytelling. 
The CD Module worked well because everyday storytellers were generally open to multiple narrative divergences as long as they remained consistent with the overall storyline logic and the defined character personas.
For other storytelling domains, the balance between human authoring and system generation must be adjusted to fit the target context. 
For example, in interactive narrative scripting for product-level games, authors are more devoted users and often require strict control over narrative logic (e.g., ensuring an event occurs only after a specific trigger~\cite{facade2005,mishra2025whatif}). 
In such contexts, while the human–AI coauthoring enabled by the CD Module may still be useful for automatically generating narrative branch options for inspiration and ideation, systems must offer more explicit control over branching structures. Authors should be able to carefully design and test each narrative path or define strict divergence conditions, rather than relying primarily on automated branching generation as in the current CD Module.


\subsection{Opportunities and Considerations for Integrating \sysname{} into Social Media}
To apply \sysname{} on real-world social media platforms, we discuss opportunities and considerations related to its interplay with conventional media, adaptation to mobile devices, and potential social interactions within the user community.

\rev{\subsubsection{Combining Interactive Vignettes with Conventional Media for Everyday Storytelling}
Building on the complementary expressive strengths of interactive vignettes and conventional media, we envision two ways of integrating these formats in everyday storytelling on social media.

First, interactive vignettes can serve as interactive annotations for text or video content. 
For example, social media users can use interactive vignettes to highlight key moments in blogs or vlogs (P12) and invite the audience to role-play, thereby adding playfulness and engagement to conventional everyday storytelling.
Second, text, images, and videos can be incorporated into interactive vignettes.
To better convey narrative progression, the interactive vignette could employ text captions that dynamically highlight words or phrases associated with the current event, helping ground players' understanding of unfolding actions~\cite{kim2021towards,kim2018facilitating}.
Beyond using text as captions, social media users could embed personal photos or video clips as storyworld assets (e.g., a family photo on the wall, a video playing on a TV), enhancing realism and grounding the interactive vignette in everyday life.
}

\subsubsection{Adaptation to Mobile Devices}
User study participants emphasized that, in everyday use, they would primarily engage with interactive vignettes on mobile devices~\cite{humphreys2013mobile}, potentially through direct sharing or through posting on existing social media platforms (e.g., X~\cite{x}, Instagram~\cite{instagram}).
The current version of \sysname{} is designed for a desktop environment, but mobile devices with smaller screens require additional design considerations for both authors and players.

For authors, the limited screen size of mobile devices cannot accommodate the three panels (environment, characters, and events) simultaneously. 
A potential design adaptation is a sequential workflow that guides authors through environment, character, and event creation step by step, with each panel displayed on a single screen and the option to revisit earlier steps.
For players, small screens may make details harder to see. 
Possible adaptations include zooming functions or a following-camera perspective on the main character to support fluid navigation within constrained display spaces.
These considerations suggest that mobile-first design will be essential for deploying \sysname{} in real-world social media contexts.

\subsubsection{Social Interactions within the User Community}
Social media is not only about publishing and viewing content but also about fostering author–player social interactions~\cite{bartolome2023literature}. 
User study participants (P8, 15) expressed curiosity about players' experiences and reactions, suggesting features such as replay logs or comment mechanisms to support two-way communication.
For instance, the author can review players' playthroughs and reply to players' comments to better communicate how the everyday stories were experienced from players' perspectives.
\rev{Additionally, to enable exploration from different perspectives, interactive vignettes can allow players to choose from multiple characters to role-play, or allow multiple players to co-participate in the same vignette.}

Among authors, we also see potential for a user community, which shares user-generated, open-sourced assets and templates for interactive vignettes, similar to user-generated Instagram AR filters~\cite{ibanez2022augmented}.
This user community could expand the diversity of visual styles, characters, and environments available for authoring. 
Incorporating such collaborative practices could expand the expressive range of interactive vignettes while aligning with familiar social media conventions.

\subsection{Ethical Considerations for AI-Generated Content in Everyday Storytelling}
Despite \sysname{} introducing the exciting opportunity to adopt interactive vignettes in everyday storytelling, we must also consider potential ethical issues that should be recognized and carefully addressed through design.
\subsubsection{Risk of Harmful or Biased Content.} Although OpenAI's usage policy helped to filter some harmful content, interactive vignettes could still be misused if authored or viewed with malicious intent. 
Characters modeled after real people may be deliberately scripted or manipulated to behave harmfully---for example, by speaking offensively, spreading misinformation, or engaging in actions that damage reputations. 
In addition, the use of LLMs introduces the risk of biased or stereotypical outputs, especially in generated conversations~\cite{weidinger2021ethical}. 
Such biases could reinforce harmful narratives or marginalize certain groups if left unchecked. We believe that deployment on social media would leverage the established moderation approaches~\cite{udanor2019combating}---including flagging, review workflows, and appeals---to ensure responsible use and safeguard users from harmful or biased content.

\subsubsection{Risk of Implanting False Memory.}
Although participants in both the formative and user studies viewed interactive vignettes not as 1:1 replicas of daily life but as interactive narratives with divergent events and dialogues, \sysname{} might risk implanting false memories~\cite{pataranutaporn2025synthetic} if players mistake AI-generated content as real interactions.
To mitigate this, we emphasize system transparency around human–AI authorship. 
The system should clearly indicate where AI-generated content is involved and prompt players to engage critically. 
This approach is similar to social media practices for deepfakes~\cite{gamage2025labeling}, where posts are labeled with warnings such as ``this post contains AI-generated content, please discern carefully.''

\section{Limitations and Future Work}
\label{sec:future}
\vspace{0.03in}
\noindent \textbf{Limitations of the User Study.}
While \sysname{} gathered in-depth qualitative feedback on the authoring and viewing experience, the user study has several limitations.
First, it was limited to one-time interactions, and a longitudinal deployment (e.g., over several weeks) would provide insights into how \sysname{} supports everyday storytelling over time (e.g., how familiarity with the system influences their authoring experience). 
Second, our participant pool was culturally homogeneous, leaving future work to explore how storytelling practices and system use may vary across cultural contexts~\cite{mcadams2019first}. 
Third, the study took place in a controlled lab setting, where the presence of researchers may have influenced participants' behaviors. 
Future work should therefore include in-the-wild, cross-cultural, and longer-term studies to better understand how \sysname{} integrates into everyday storytelling practices.
\rev{To enable long-term, repeated use of \sysname{}, the system would need features that let users reuse environment templates and characters.}

\rev{
\vspace{0.03in}
\noindent \textbf{Extending \sysname{}'s Story Scope.}
The current version of \sysname{} focuses on simple, event-centric everyday stories; future extensions could broaden its story scope to include more characters, transitions across multiple locations, and richer emotional expression.
Supporting stories with more characters and transitions between spatially discontinuous locations (e.g., moving from home to school) would allow \sysname{} to represent more complicated and richer slices of everyday life. 
Enhancing the system's ability to capture emotional expression would further expand its expressive range, which requires extending the interactive vignette specification to include characters' emotional description, their evolution across the events, and their influence on character behaviors over time.
However, these extensions increase authoring complexity, as users must manage a larger set of narrative elements. Future systems should therefore explore intelligent interface support, such as LLM-assisted timeline-based authoring tools~\cite{mishra2025whatif}, to keep authoring effort manageable while still enabling flexible editing across multiple story components.

\vspace{0.03in}
\noindent \textbf{Extending \sysname{}' Viewing Experience.}
While participants appreciated the current viewing experience of interactive vignettes in \sysname{}, future work can extend its social aspects and spatial expressiveness: allowing players to choose different characters, supporting multiplayer co-participation, and expanding from 2D to 3D spatial storytelling.
Supporting character selection would require authors to prepare core narrative paths for each character, ensuring that switching characters provides meaningfully different role-play perspectives. 
Enabling multiplayer co-participation introduces a different challenge: the system will need to coordinate multiple players while still maintaining the author's intended storyline, rather than drifting into a fully open-ended narrative.
At the same time, expanding to 3D storytelling poses significant challenges. 
Although recent text-to-3D advances (e.g., Genie 3~\cite{parkerholder2025genie3}, SceneTeller~\cite{ocal2024sceneteller}, LayoutGPT~\cite{feng2023layoutgpt}) demonstrate the feasibility of generating complex 3D worlds from narrative text, systems must still ensure that generated scenes include all story-relevant objects and relations, and remain easily editable so authors can refine layouts and maintain creative control. 
A 3D environment may also raise expectations for realistic character behaviors (e.g., cooking, eating, jumping), further increasing authoring effort. 
Realizing these extensions will therefore require future systems to provide effective generation supports and lightweight refinement tools to avoid overwhelming everyday storytellers.
}

\vspace{0.03in}
\noindent \textbf{Potential Applications Beyond Everyday Storytelling.}
While \sysname{} was designed to support everyday storytelling, its low-effort authoring process, roleplay mechanics, and the branch-and-bottleneck narrative structure suggest potential for broader applications.
First, in the interactive media industry~\cite{engstrom2018prototyping}, \sysname{} could serve as a rapid prototyping tool for designers to test and communicate ideas visually.
Second, in mental health counseling~\cite{ronning2019use}, interactive vignettes might be customized to support ad hoc interactive narratives, though such use would require careful ethical and clinical consideration.
Third, in education~\cite{spierling2006learning}, adapting the system for age-appropriate, curriculum-aligned content could position interactive vignettes as an engaging learning tool.
Extending \sysname{} beyond everyday storytelling would require rethinking the vignette specification, integrating domain-specific constraints, and aligning with professional practices. 
Collaboration with domain experts, such as game designers, clinicians, and educators, will be essential to guide appropriate adaptation and evaluation. 
These opportunities point toward promising directions for future research and development.

\section{Conclusion}
In this work, we introduced \sysname{}, an AI-assisted authoring system that enables lightweight creation of interactive vignettes for everyday storytelling. 
Interactive vignettes allow the audience to role-play the author's perspective and actively participate in the story, which motivated social media users to adopt the format into everyday storytelling.
However, authoring interactive vignettes using existing authoring tools is technically demanding and time-intensive, which conflicts with the immediacy required in everyday storytelling. 
At the same time, the aspiration to provide branching narratives with adaptive NPC behaviors that allow divergent activities beyond the core storyline further increases authoring complexity.
\sysname{} addresses these challenges by automatically transforming text stories into interactive vignette elements and guiding authors in refinement, and using the CD Module to automatically plan NPC behaviors that adapt to player interactions while preserving the author's intended story message. 
A preliminary technical evaluation showed that the module generates believable NPC activities based on the author-defined character personas and the overarching storyline. 
A user study demonstrated that \sysname{} supports low-effort authoring while enabling engaging viewing experiences that convey the core story message.
Finally, we discuss the real-world applicability of \sysname{} on social media platforms and highlight potential extensions beyond everyday storytelling. 
\rev{Overall, while our research is among the early efforts toward enabling interactive vignette creation in everyday use with several limitations, we believe that the system hints at a promising new form of social everyday storytelling and offers insights and agendas for future work.}


\begin{acks}
\textcolor{black}{
This work was supported by a grant of the KAIST-KT joint research project funded by KT (G01220645), the Institute of Information \& Communications Technology Planning \& Evaluation (IITP) grant funded by the Korean government (MSIT) (No. 2021-0-01347, Video Interaction Technologies Using Object-Oriented Video Modeling), the Culture, Sports and Tourism R\&D Program through the Korea Creative Content Agency grant funded by the Ministry of Culture, Sports and Tourism in 2026 (Project Name: Development of Multimodal UX Evaluation Platform Fechnology for XR Spatial Responsive Content Optimization, Grant No.: RS-2024-00361757, Contribution Rate: 20\%), and the 2025-1 Yonsei University Future-Leading Research Initiative (Grant No.: 2025-22-0156).}
We used ChatGPT~\cite{chatgpt} to perform grammar checks, as well as obtain phrase suggestions and grammatical restructuring based on our own texts.
\end{acks}

\bibliographystyle{ACM-Reference-Format}
\bibliography{reference}

\appendix

\appendix

\section{Participants' Background Information}
\label{app:participant}
\onecolumn
\begin{table*}
\newcommand{\thinrule}{\specialrule{0.02pt}{0pt}{0pt}}
\renewcommand{\arraystretch}{1.3}
\footnotesize
\centering
\caption{Pilot study participants' background information. Gender is abbreviated as M = Male and F = Female. Weekly Frequency is abbreviated as Wk. Freq. Experience is abbreviated as Exp.}
\begin{tabularx}{\textwidth}{
    p{0.3cm}   
    p{0.6cm}   
    p{0.3cm}   
    p{2.8cm}   
    X          
    X          
    X          
    X          
}
\toprule
\textbf{ID} &
\textbf{Gender} &
\textbf{Age} &
\textbf{Major} &
\multicolumn{2}{c}{\textbf{Everyday Storytelling}} &
\multicolumn{2}{c}{\textbf{Interactive Vignettes}} \\
\cmidrule(lr){5-6} \cmidrule(lr){7-8}
& & & &
\textbf{Viewing Wk. Freq.} &
\textbf{Authoring~Wk.~Freq.} &
\textbf{Playing Wk. Freq.} &
\textbf{Authoring Exp.} \\
\midrule
P1  & M & 21 & Brain Science               & \textgreater5 & 3--5 & 1--2 & None \\
\thinrule
P2  & M & 26 & Electrical Engineering      & \textgreater5 & 1--2 & \textgreater5 & Built storyworlds in Minecraft \\
\thinrule
P3  & F & 23 & Culture Technology          & \textgreater5 & 3--5 & 1--2 & None \\
\thinrule
P4  & M & 35 & Material                    & \textgreater5 & 3--5 & 3--5 & None \\
\thinrule
P5  & F & 25 & Industrial Design           & \textgreater5 & \textgreater5 & \textgreater5 & Tried to make a 3D character in Unity \\
\thinrule
P6  & F & 27 & Chemistry                   & 3--5 & \textgreater5 & 1--2 & None \\
\thinrule
P7  & F & 22 & Biomedical Sciences         & \textgreater5 & \textgreater5 & 3--5 & None \\
\thinrule
P8  & M & 21 & Computer Science            & \textgreater5 & 3--5 & 0 & None \\
\thinrule
P9  & M & 32 & Computer Science            & 1--2 & \textgreater5 & 3--5 & None \\
\thinrule
P10 & F & 21 & Education                   & 3--5 & 1--2 & 3--5 & None \\
\thinrule
P11 & M & 19 & Robotics                    & 3--5 & 3--5 & \textgreater5 & None \\
\thinrule
P12 & M & 24 & Transportation Engineering  & \textgreater5 & \textgreater5 & 1--2 & None \\
\thinrule
P13 & F & 23 & Electrical Engineering      & \textgreater5 & 3--5 & 3--5 & None \\
\bottomrule
\end{tabularx}
\label{tab:pilotparticipants}
\end{table*}

\begin{table*}
\footnotesize
\renewcommand{\arraystretch}{1.1}
\centering
\caption{User study participants' background information. Gender is abbreviated as M = Male and F = Female. Icons in the last column indicate experience categories: \faPenNib{} = Text Generation, \faImage{} = Image Creation, \faVideo{} = Video Creation, and \faCode{} = Coding. Weekly Frequency is abbreviated as Wk. Freq. Experience is abbreviated as Exp.}
\begin{tabularx}{\textwidth}{
    p{0.3cm}
    p{0.6cm}
    p{0.3cm}
    p{2.6cm}
    X
    X
    X
    X
    p{2.2cm}
}
\toprule
\textbf{ID} &
\textbf{Gender} &
\textbf{Age} &
\textbf{Major} &
\multicolumn{2}{c}{\textbf{Everyday Storytelling}} &
\multicolumn{2}{c}{\textbf{Interactive Vignettes}} &
\textbf{AI-Assisted Authoring Exp.} \\
\cmidrule(lr){5-6} \cmidrule(lr){7-8}
& & & &
\textbf{Viewing Wk. Freq.} &
\textbf{Authoring Wk. Freq.} &
\textbf{Playing Wk. Freq.} &
\textbf{Authoring Exp.} &
\\
\midrule
P1  & M & 22 & Aerospace Engineering & \textgreater5 & \textgreater5 & 1--2 & None & \faPenNib{} \\
\midrule
P2  & M & 26 & Cognitive Sciences & 3--5 & 3--5 & 1--2 & None & \faPenNib{}, \faImage{} \\
\midrule
P3  & M & 21 & Computer Science & \textgreater5 & 3--5 & \textgreater5 & None & \faPenNib{}, \faCode{} \\
\midrule
P4  & F & 21 & Business and Technology Management & \textgreater5 & 3--5 & 3--5 & None & \faPenNib{} \\
\midrule
P5  & M & 21 & Industrial Design & 3--5 & 1--2 & \textgreater5 & None & \faPenNib{}, \faVideo{} \\
\midrule
P6  & M & 26 & Culture Technology & 3--5 & \textgreater5 & 1--2 & None & \faPenNib{} \\
\midrule
P7  & M & 22 & Bio and Brain Engineering & \textgreater5 & 3--5 & 3--5 & None & \faPenNib{}, \faImage{} \\
\midrule
P8  & M & 22 & Physics & \textgreater5 & \textgreater5 & 0 & None & \faPenNib{} \\
\midrule
P9  & F & 25 & Business and Technology Management & 1--2 & \textgreater5 & 3--5 & None & \faPenNib{}, \faImage{} \\
\midrule
P10 & M & 21 & Computer Science & 3--5 & 3--5 & 3--5 & None & \faPenNib{},\faCode{} \\
\midrule
P11 & F & 19 & Undeclared Major & \textgreater5 & \textgreater5 & \textgreater5 & Built an RPG using Unity & \faPenNib{} \\
\midrule
P12 & M & 23 & Computer Science & \textgreater5 & 1--2 & 1--2 & None & \faPenNib{}, \faCode{} \\
\midrule
P13 & M & 18 & Undeclared Major & \textgreater5 & 3--5 & 3--5 & None & None \\
\midrule
P14 & M & 23 & Physics & \textgreater5 & \textgreater5 & 1--2 & None & \faImage{} \\
\midrule
P15 & M & 22 & Computer Science & 3--5 & 1--2 & 0 & None & \faImage{}, \faCode{} \\
\midrule
P16 & M & 23 & Computer Science & \textgreater5 & 3--5 & 1--2 & None & \faCode{} \\
\bottomrule
\end{tabularx}
\label{tab:userstudyparticipants}
\end{table*}


\end{document}